%% file: preprint.tex
\documentclass[runningheads,11pt]{llncs}
\usepackage[top=2.54cm, bottom=2.54cm, left=2.54cm, right=2.54cm]{geometry}
\input{macros}
\graphicspath{{figs/}}
\begin{document}

\title{Nonlinear ensemble filtering with diffusion models:\\ Application to the surface quasi-geostrophic dynamics}

\author{{Feng Bao}\inst{1}\href{https://orcid.org/0000-0002-1302-8120}{\includegraphics[scale=0.08]{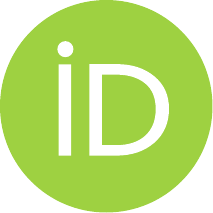}} \and Hristo G.~Chipilski\thanks{\footnotesize \textit{Corresponding author}: \textit{Dr.~Hristo G.~Chipilski,} \texttt{hchipilski@fsu.edu}.\\[1mm]
This Work has been submitted to Monthly Weather Review. Copyright in this Work may be transferred without further notice.}\inst{2} \href{https://orcid.org/0000-0003-3287-0038}{\includegraphics[scale=0.08]{orcid.pdf}} \and Siming Liang\inst{1} \and \\ Guannan Zhang\inst{3} \and Jeffrey S.~Whitaker\inst{4}}

\institute{Department of Mathematics, Florida State University, Tallahassee, Florida \and Department of Scientific Computing, Florida State University, Tallahassee, Florida \and Computer Science and Mathematics Division, Oak Ridge National Laboratory \and Physical Sciences Laboratory, NOAA/Earth System Research Laboratories, Boulder, Colorado}
\maketitle

% headers
\pagestyle{fancy}
\fancyhf{}
\fancyhead[L]{\textit{Nonlinear ensemble filtering with diffusion models}}
\fancyhead[R]{\textit{Bao et al.}}
\fancyfoot[C]{\thepage}

% display fist page number
\thispagestyle{plain}

%-=-=-=-=-=-=-=-=-=-=-=-=-=-=-=-=-=-=
%           ABSTRACT
%-=-=-=-=-=-=-=-=-=-=-=-=-=-=-=-=-=-=
\begin{abstract}
The intersection between classical data assimilation methods and novel machine learning techniques has attracted significant interest in recent years. Here we explore another promising solution in which diffusion models are used to formulate a robust nonlinear ensemble filter for sequential data assimilation. Unlike standard machine learning methods, the proposed \textit{Ensemble Score Filter (EnSF)} is completely training-free and can efficiently generate a set of analysis ensemble members. In this study, we apply the EnSF to a surface quasi-geostrophic model and compare its performance against the popular Local Ensemble Transform Kalman Filter (LETKF), which makes Gaussian assumptions on the posterior distribution. Numerical tests demonstrate that EnSF maintains stable performance in the absence of localization and for a variety of experimental settings. We find that EnSF achieves competitive performance relative to LETKF in the case of linear observations, but leads to significant advantages when the state is nonlinearly observed and the numerical model is subject to unexpected shocks. A spectral decomposition of the analysis results shows that the largest improvements over LETKF occur at large scales (small wavenumbers) where LETKF lacks sufficient ensemble spread. Overall, this initial application of EnSF to a geophysical model of intermediate complexity is very encouraging, and motivates further developments of the algorithm for more realistic problems.

\keywords{diffusion models \and ensemble data assimilation \and surface quasi-geostrophic turbulence}
\end{abstract}

%-=-=-=-=-=-=-=-=-=-=-=-=-=-=-=-=-=-=
%           INTRODUCTION
%-=-=-=-=-=-=-=-=-=-=-=-=-=-=-=-=-=-=
\section{Introduction}

Since its introduction by \citet{evensen_1994}, the ensemble Kalman filter (EnKF) has been utilized extensively for the initialization of geophysical models and has inspired the rapidly developing subfield of ensemble data assimilation (DA). In numerical weather prediction (NWP), the ensemble approach has been highly successful either on its own or in conjunction with variational methods such as 4D-Var \citep[e.g.,][]{isaksen_et_al_2010}. This success largely stems from the ability of a forecast ensemble to capture the ``errors of the day'', replacing the assumptions of static forecast covariances used in earlier DA methods (e.g., optimal interpolation) and purely variational approaches. 

In meteorological applications, EnKF algorithms have been particularly beneficial at the convective scales due to their ability to handle complex numerical models and observing systems \citep[e.g., ][]{aksoy_et_al_2009,aksoy_et_al_2010,jones_et_al_2016,chipilski_et_al_2020,chipilski_et_al_2022,hu_et_al_2023}. One of the practical advantages of ensemble DA methods is their ability to incorporate highly nonlinear models and observations without the need to explicitly compute any tangent linear and adjoint operators. This is an extra overhead for any variational DA method in operational contexts as one needs to continuously recompute these operators once a new version of the model is released. Nevertheless, standard EnKFs still approximate the Kalman filter's analysis equations, which are derived from restrictive Gaussian assumptions. These Gaussian assumptions manifest themselves in the linear nature of the EnKF's update, which limits the ability to represent more complex analysis (posterior) distributions \citep{spantini_et_al_2022}. Recent studies have shown that the analysis biases introduced by methods leveraging the Kalman filter equations, such as the EnKF, can have a detrimental impact on the ability to forecast high-impact weather events like hurricanes \citep{poterjoy_2022}.

Particle filters \citep[PFs;][]{gordon_et_al_1993,vanLeeuwen_2009,vanLeeuwen_et_al_2019} represent a natural replacement for standard EnKFs due to their provable convergence to the correct posterior distribution \citep{crisan_doucet_2002}. Although they were introduced around the same time as EnKFs, they have taken a long time to reach operational potential. The underlying reasons have practical dimensions -- in order to avoid the curse of dimensionality, PFs require a prohibitively large number of particles. There has been visible progress in this direction, with several great examples of successful PF implementations in large systems \citep{todter_et_al_2016,poterjoy_et_al_2017,rojahn_et_al_2023}. PFs have also stimulated the development of many complementary non-Gaussian DA approaches, including lognormal and bi-Gaussian extensions of standard DA methods \citep{fletcher_2010,fletcher_et_al_2023,chan_et_al_2020} and the recently developed two-step quantile-conserving ensemble filtering framework (QCEFF) of \citet{anderson_2022,anderson_2023}.

It is clear that research into nonlinear/non-Gaussian ensemble DA methods will continue to attract more interest in view of the increasing complexity of numerical models and observing systems. One promising way to address these challenges is to exploit the ongoing revolution in generative artificial intelligence \citep[GenAI; see][]{DBLP:conf/iccv/LuoH21,song2021scorebased, baranchuk2022labelefficient}. Up to now, two GenAI approaches have been adopted in the ensemble DA context. \citet{chipilski_2023} showed how the appealing mathematical properties of invertible neural networks (normalizing flows) can help generalize the Kalman Filter to arbitrarily non-Gaussian distributions. The resulting Conjugate Transform Filter (CTF) is amenable to ensemble approximations which can take advantage of existing EnKF solvers. The second GenAI framework is associated with the score-based filter of \citet{SF_2023} which harnesses the expressive power of diffusion models to approximate complex posterior distributions. The use of a pseudo time to gradually transform samples to the desired distribution makes this approach somewhat similar to the Particle Flow Filter \citep[PFF; ][]{pulido_vanLeeuwen_2019,hu_vanLeeuwen_2021}, but its ability to incorporate non-Gaussian priors is an important distinction. Following its original formulation, a more efficient ensemble version of the score-based filter, henceforth referred to as the \textit{Ensemble Score Filter (EnSF)}, has been developed. Owing to its training-free nature, EnSF has been shown to estimate the states of a high-dimensional Lorenz-96 system with up to $10^6$ variables \citep{EnSF_2023}. More recently, the EnSF framework has been also extended to conduct joint state-parameter estimation \citep{UF_2023}.

In this study, we focus on the further development of the EnSF algorithm and aim to introduce its capabilities to the geophysical community. Motivated by its scalable performance in high-dimensional and nonlinear settings, we implement EnSF in a surface quasi-geostrophic (SQG) model which provides a realistic description of geophysical turbulence \citep{smith_et_al_2023}. Our main objective is to demonstrate that (i) EnSF maintains stable performance in high dimensions while (ii) offering advantages compared to standard EnKF methods. This is achieved by performing a hierarchy of SQG experiments in which we gradually increase the complexity of the DA task toward more nonlinear settings.

The rest of this paper is organized as follows. In Section \ref{sec:methodology}, we introduce basic theory from diffusion models and discuss how the associated backward (reverse-time) stochastic differential equations (SDEs) can be used to sample from the Baysian posterior. After describing important experimental design choices in Section \ref{sec:experimental_design}, we present our DA results with the SQG model in Section \ref{sec:results}. In Section \ref{sec:conclusions}, we conclude with a summary of all main findings and outline several interesting research directions.

%-=-=-=-=-=-=-=-=-=-=-=-=-=-=-=-=-=-=
%           METHODOLOGY
%-=-=-=-=-=-=-=-=-=-=-=-=-=-=-=-=-=-=
\section{Methodology} \label{sec:methodology}

We first introduce the diffusion model as a general framework to sample from a user-specified target distribution. Then, we explain how score-based diffusion models fit the standard forecast-analysis procedure of sequential DA methods. Finally, we discuss numerical schemes for solving the DA problem under the score-based filtering framework, and provide details on how to build ensemble approximations.

\subsection{Introduction to diffusion models}
In a diffusion model, there is a forward $\sR^{N_x}$-valued stochastic differential equation (SDE) defined as
\begin{equation}\label{forward:SDE}
d \rv{Z}_t = b(t)\rv{Z}_t dt + \sigma(t)d\rv{W}_t \quad \text{(forward SDE),}
\end{equation}
where $\rv{W}_t$ is a standard $\sR^{N_x}$-valued Brownian motion (Wiener process) corresponding to an It\^o type stochastic integral $\int \cdot d\rv{W}_t$, while $b$ and $\sigma$ are two explicitly given functions referred to as the drift and diffusion coefficients, respectively. The initial condition $\rv{Z}_0$ of the SDE in Eq.~\eqref{forward:SDE} follows some target distribution with its probability density function (pdf) denoted by $Q(\bz_0)$. One can show that with properly chosen $b$ and $\sigma$, the diffusion process $\{\rv Z_t\}_{0\leq t \leq T}$ can transform any target pdf ${Q}(\bz_0)$ to a standard Gaussian, i.e. $\rv Z_T \sim \mathcal{N}(\bo, \bI)$, where $[0, T]$ is often referred to as the \textit{pseudo-time interval}. 

Generating samples $\{\rv Z_0^i \}_{i=1}^{N}$ of the target random variable $\rv Z_0$ pertains to simulating the following reverse-time SDE:
\begin{equation}\label{reverse:SDE}
\begin{aligned}
d\rv Z_t = \left[ b(t)\rv Z_t -\sigma^2(t) \bs(\rv Z_t,t) \right] dt &+ \sigma(t)d\overleftarrow{\rv W}_t \quad \text{(reverse-time SDE),}
\end{aligned}
\end{equation}
where $\int \cdot d\overleftarrow{\rv W}_t$ is a \textit{backward} It\^o stochastic integral \citep{SDE, BDSDE}, and $\bs(\cdot, t)$ is the so-called \textit{score function} given by

\begin{equation}\label{eq:score_func}
\bs(\bz_t, t) \coloneqq \nabla \log Q(\bz_t),
\end{equation}
where $Q(\bz_t)$ denotes the pdf of $\rv Z_t$.
Note that the reverse-time SDE in Eq.~\eqref{reverse:SDE} is also a diffusion process except that the propagation direction is backwards in time from $T$ to $0$ with initial condition $\rv Z_T$ given at time $T$. An important result from the literature is that the solution $\rv Z_0$ of the reverse-time SDE follows the target distribution \citep{SF_2023}. The practical implications are that we can generate samples from a standard Gaussian distribution (which can be done efficiently) and use the reverse-time SDE to transform them to samples of the target distribution. The score function $\bs(\cdot, t)$ has an important role in this mapping process as it stores information about the distribution of the samples, which in turn helps guide their transformation over the pseudo-time interval as $T \rightarrow 0$. In particular, having the score function associated with the target pdf $Q(\bz_0)$ and the predefined forward SDE allows us to generate an unlimited number of target samples by running the reverse-time SDE in Eq.~\eqref{reverse:SDE}.

In both the forward SDE and the reverse-time SDE, one needs to carefully choose the drift and diffusion coefficients $b$ and $\sigma$. While there are multiple options for $b$ and $\sigma$, in this work we let

\begin{equation}\label{coefficients}
\begin{aligned}
    &b(t) = \frac{d\log \alpha_{t}}{dt}\\[4mm]
&\sigma^2(t) = \frac{d \beta^2_{t}}{d t} - 2 \frac{d \log \alpha_{t}}{d t} \beta^2_{t}
\end{aligned}
\end{equation}
\vspace*{2mm} with $\alpha_{t} = 1 - t$ and $\beta_{t} = \sqrt{t}$ for $t \in [0, 1]$, which is consistent with the choice made in \citet{song2021scorebased}. Nevertheless, in future work we plan to explore the sensitivity of EnSF to different options for the drift and diffusion coefficients. 

The traditional use of diffusion models in the machine learning (ML) literature is to generate highly realistic images and videos \citep{song2021scorebased}. Later in our exposition, we will discuss how this powerful approach can be also used in the context of data assimilation (DA).

\subsection{The general filtering solution}

The formulation of every DA method requires two main ingredients -- (stochastic) dynamic and observation models. Let $k = 0, 1, 2, ..., K$ be a time index. A general representation of the discretized system evolution is given by 

\begin{equation}\label{state}
\rv X_k = \bvf_{k-1}(\rv X_{k-1}, \rv E^m_{k-1}),
\end{equation}

\noindent where $\rv X_k \in \sR^{N_x}$ stands for the (true) state vector we wish to estimate and $\bvf_k$ represents the error-prone numerical model. The model errors $\rv E^m_k$ arise due to (i) discretization of the governing equations and (ii) imperfect knowledge about the simulated process. In this study, we assume the model errors are known and we only need to infer the unknown true state from the noisy observations

\begin{equation}\label{observations}
\rv Y_{k} = \bvh_k(\rv X_k) + \rv E^o_k \quad \text{with } \rv E^o_k \sim \mathcal{N}(\bo,\bR_k),
\end{equation}

\noindent where $\bvh_k$ is an observation operator and $\rv E^o_k$ are the associated observation errors. Note that while the formulation of Eq.~\eqref{observations} follows the standard additive-Gaussian observation error model, it is possible to introduce more general statistical assumptions \citep{chipilski_2023}.  

Given the model and observation equations, a rather general formulation of the DA problem is to seek the filtering (posterior) pdf $P(\bx_k|\by_{1:k})$\footnote{The notation $1:k$ is shorthand for the set of integers from $1$ to $k$.}, which can be done recursively by iterating through a prediction and an update step:

\paragraph*{Prediction:}
Given the filtering pdf $P(\bx_{k-1} | \by_{1:k-1})$ at the previous time level $k-1$, we can first compute the prior pdf $P(\bx_k | \by_{1:k-1})$ using the Chapman-Kolmogorov equation

\begin{equation}\label{Kolmogorov}
P(\bx_k | \by_{1:k-1}) = \int P(\bx_{k-1} | \by_{1:k-1}) P(\bx_k | \bx_{k-1}) d\rv \bx_{k-1},
\end{equation}

\noindent where $P(\bx_k | \bx_{k-1})$ is the transition pdf determined from the dynamic model in Eq.~\eqref{state}.

\paragraph*{Update:} 
After receiving the new set of observations $\rv Y_k = \by_k$, we can adjust the forecast state towards the observations by applying Bayes' theorem,
\begin{equation}\label{Bayesian}
P(\bx_k | \by_{1:k}) \propto P(\bx_k | \by_{1:k-1}) P(\by_k|\bx_k),
\end{equation}
where the likelihood function $P(\by_k|\bx_k)$ is determined from the observation model in Eq.~\eqref{observations}. In view of our Gaussian assumptions, the likelihood can be written as

\begin{equation}\label{Likelihood}
\begin{aligned}
    P(\by_k|\bx_k) \propto 
     \exp\Big( - \frac{1}{2}  [\by_k- \bvh(\bx_k)]\T \bR_k^{-1} [\by_k- \bvh(\bx_k)] \Big).
\end{aligned}
\end{equation}

\noindent Through the recursive application of the prediction and update steps in Eqs.~\eqref{Kolmogorov} and \eqref{Bayesian}, we can find the filtering pdf for any time index $k$. 

\subsection{Score-based filtering for data assimilation}
The central idea in score-based filtering is to create score models which represent the filtering pdfs at different times and then generate analysis ensemble members according to the reverse-time SDE in Eq.~\eqref{reverse:SDE}. To this end, we introduce the score functions $\bs_{k | k-1}$ and $\bs_{k | k}$ corresponding to the prior pdf $P(\bx_k | \by_{1:k-1})$ and the posterior pdf $P(\bx_k | \by_{1:k})$:

\begin{itemize}
\item Prior filtering score $\bs_{k | k-1}$ such that $\rv Z_0 \equiv \rv X_k|\rv Y_{1:k-1}$; i.e., $Q(\bz_0) \equiv P(\bx_k | \by_{1:k-1})$.
\sT \item Posterior filtering score $\bs_{k | k}$ such that $\rv Z_0 \equiv \rv X_k|\rv Y_{1:k}$; i.e., $Q(\bz_0) \equiv P(\bx_k | \by_{1:k})$.
\end{itemize}

\noindent Next, we describe a recursive procedure to implement the score-based filter for DA. To proceed, it is assumed that the posterior score $\bs_{k-1|k-1}$ associated with the posterior (filtering) pdf at time level $k-1$ is given. This is analogous to how the initial state distribution is assumed to be known in sequential DA. An Euler scheme can be used to discretize the reverse-time SDE in Eq.~\eqref{reverse:SDE} \citep{SDE} as follows

\begin{equation}\label{Scheme:reverse}
\begin{aligned}
\rv Z^m_{t_n} = \rv Z^m_{t_{n+1}} - \big[b(t_{n+1}) \rv Z^m_{t_{n+1}}& -\sigma^2(t_{n+1}) \bs_{k-1|k-1}(\rv Z^m_{t_{n+1}}, t_{n+1})\big] \Delta t_n - \sigma(t_{n+1}) \Delta \rv W_{t_n},\\
& n = 0, 1, \dots, N-1; \quad m = 1, 2, \dots, M,
\end{aligned}
\end{equation}

\noindent where $\{ \rv Z^m_{t_n} \}_{m=1}^M$ is a set of $M$ iid samples, $\Delta t_n \coloneqq t_{n+1} - t_n$, $\Delta \rv W_{t_n} \coloneqq \rv W_{t_{n+1}} - \rv W_{t_n}$, and the above scheme uses the following time discretization 

$$\{t_n: 0 = t_0 < t_1 < \cdots < t_n < \cdots < t_N = T\}$$

\noindent over the diffusion model's pseudo-time interval $[0, T]$. The Euler scheme is one of the two most popular schemes for solving SDEs numerically. The second one is based on the Milstein method \citep{milstein_1975}, but for the standard SDE formulation used here (in which the diffusion coefficient $\sigma$ is state independent), the two schemes give identical results.

Following the standard workflow of diffusion models, the reverse-time SDE is initialized with samples drawn from a standard Gaussian distribution; that is, $\{\rv Z^m_{t_{N}}\}_{m=1}^M \sim \mathcal{N}(\bo, \bI)$. Leveraging the Euler scheme in Eq.~\eqref{Scheme:reverse}, these initial samples are mapped to the analysis ensemble $\{\rv X^{m}_{k-1|k-1}\}_{m=1}^{M}$ which follows the filtering pdf $P(\bx_{k-1} | \by_{1:k-1})$ by construction.

The generative nature of the diffusion process needed to obtain the analysis ensemble $\{\rv X^{m}_{k-1|k-1}\}_{m=1}^{M}$ is worth elaborating on. With an appropriately estimated score function $\hat{\bs}_{k-1|k-1}$, we can generate unlimited samples from the posterior pdf $P(\bx_{k-1} | \by_{1:k-1})$, which can be seen as a non-Gaussian extension of the class of resampling EnKFs discussed by \citet{anderson_2001}. In addition, the sample size $M$ can be an arbitrary number depending on the specific application and computing resources.  

\paragraph*{Prediction step:} The prediction step in the score filter is analogous to all ensemble DA methods. In particular, the dynamic model in Eq.~\eqref{state} is used to advance each analysis ensemble member $\rv X^{m}_{k-1|k-1}$ to the next observation time level $k$. In doing so, we obtain the forecast ensemble $\{\rv X^{m}_{k|k-1} \}_{m=1}^M$ which approximates the prior pdf $P(\bx_k | \by_{1:k-1})$ and will be used to estimate the prior score function $\hat{\bs}_{k| k-1}$. 

\paragraph*{Update step:} The posterior score function $\bs_{k | k}$ implicitly encodes the posterior pdf $P(\bx_k|\by_{1:k})$, which is why its estimation is an essential part of the update step in the score filter. Since $P(\bx_k|\by_{1:k})$ combines information from both the prior and the likelihood, the derivation of $\bs_{k | k}$ will focus on how to effectively adjust the prior score based on the assimilated observations. Indeed, this is a general challenge for all DA methods. Calculating the gradient of the logarithm of Eq.~\eqref{Bayesian}, we see that

\begin{equation}
\begin{aligned}
 \nabla_\bx \log P(\bx_{k} | \by_{1:k}) = \nabla_\bx \log P(\bx_k | \by_{1:k-1}) +  \nabla_\bx \log P(\by_k | \bx_k),
\end{aligned}
\end{equation}
where the gradient $\nabla_\bx $ is taken with respect to the state variable $\bx_k$.

The posterior pdf $P(\bx_k | \by_{1:k})$ is the target distribution for the update step of the score filter, implying that $\nabla_\bx \log P(\bx_k | \by_{1:k})$ is the desired posterior score function $\bs_{k|k}$ at pseudo-time $t=0$ (recall the score function definition in Eq.~\ref{eq:score_func}). Further notice that $P(\bx_k | \by_{1:k-1})$ is the prior pdf, hence $\nabla_\bx \log P(\bx_k | \by_{1:k-1})$ is equivalent to the prior score function $\bs_{k|k-1}$ at pseudo time $t=0$. This suggests that $\nabla_\bx \log P(\by_k|\bx_k)$ should be the likelihood portion of the posterior filtering score at pseudo time $t=0$. Taking advantage of this link, we propose the following posterior score function

\begin{equation}\label{S-posterior}
\bs_{k | k}(\bz, t) := \bs_{k|k-1}(\bz, t) + h(t) \nabla_\bx \log p(\by_k | \bz).
\end{equation}
The prior score $\hat{\bs}_{k | k-1}$ is already estimated from the forecast ensemble, and we will refer to $\nabla_\bx \log p(\by_k |\cdot)$ as the likelihood score associated with the observational data $\by_k$. The coefficient $h(t)$ multiplying the likelihood score is a \textit{damping function} satisfying the following property: 

\begin{center}
  \item  $h(t)$ is monotonically decreasing in $[0,T]$ with $h(0) = 1$ and $h(T) = 0$.
\end{center}
We use $h(t) = T - t$ for the SQG experiments reported in this paper. However, it should be emphasized there are multiple choices to define $h$, and the question of which one is mathematically optimal remains open.

Having the posterior score $\hat{\bs}_{k|k}$, we can now use the discretized scheme in Eq.~\eqref{Scheme:reverse} to transport a set of samples from $\mathcal{N}(\bo,\bI)$ to the desired analysis ensemble $\{\rv X_{k|k}^m\}_{m=1}^M$ from $P(\bx_k | \by_{1:k})$. 

\subsection{The Ensemble Score Filter (EnSF)}
The estimation of score functions is a central topic in the score-based filtering approach discussed so far. The standard technique to estimate scores is via deep learning \citep{song2021scorebased, SF_2023}. Here, we introduce an ensemble approximation referred to as the \textit{Ensemble Score Filter (EnSF)} which does not require the training of neural networks.

Since the likelihood score appearing in Eq.~\eqref{S-posterior} can be computed explicitly in the Gaussian case considered here, the major computational effort in EnSF lies in evaluating the prior score $\hat{\bs}_{k|k-1}$. The latter can be achieved by setting $\rv Z_0$ to the forecast ensemble $\{\rv X^{m}_{k|k-1}\}_{m=1}^{M}$. From the definition of score functions and the choice of $b$ and $\sigma$ in Eq.~\eqref{coefficients}, we have that the conditional density $Q(\bz_t|\bz_0)$ needed in the forward SDE is given by

\begin{equation}
Q(\bz_t|\bz_0) \propto \exp\Big( - \frac{1}{2 \beta_t^2}  (\bz_t - \alpha_t \bz_0)\T (\bz_t - \alpha_t \bz_0) \Big).
\end{equation}

\s \sO \noindent With this, marginalizing over $\bz_0$ yields the following score function

\s \begin{equation}\label{eq:score}
\begin{aligned}
\bs(\bz_{t}, t) & = \nabla_\bz \log Q(\bz_t) = \nabla_\bz \log \left(\int Q({\bz}_t | \bz_0) Q(
\bz_0) d\bz_0\right)\\[2mm]
& = \frac{1}{\int Q(\bz_t | \bz'_0) Q(
\bz'_0) d\bz'_0}  \int  - \frac{\bz_t - \alpha_t \bz_0}{\beta^2_t} Q(\bz_t | \bz_0) Q(\bz_0) d\bz_0\\[2mm]
& =  - \int \frac{\bz_t- \alpha_t \bz_0}{\beta^2_t} \bw_t(\bz_t,  \bz_0)  Q(\bz_0)d\bz_0,
\end{aligned}
\end{equation}

\s \s \noindent where the weight function $\bw_t(\bz_t,  \bz_0)$ is defined by

\s \begin{equation}\label{eq:weight}
\bw_t(\bz_t,\bz_0) :=  \frac{ Q(\bz_t | \bz_0) }{\int Q(\bz_t | \bz'_0) Q(\bz'_0) d\bz'_0},
\end{equation}

\s \sO \noindent and satisfies the condition $\int \bw_t(\bz_t,  \bz_0) Q(\bz_0) d\bz_0 = 1$. 

Then, we can apply a Monte Carlo approximation for $\bs_{k | k-1}$ based on Eq.~\eqref{eq:score} at a given $\bz$ and $t \in [0, 1]$:

\s \begin{equation}\label{Approx:S-prior}
\bs_{k|k-1}(\bz, t) \approx \hat{\bs}_{k|k-1}(\bz, t) := \sum_{j=1}^{J} - \frac{\bz - \alpha_{t} \bx_{k|k-1}^{m_j}}{\beta^2_{t}} \hat{\bw}_{t}\left(\bz, \bx_{k|k-1}^{m_j} \right),
\end{equation}

\s \noindent where $\{\rv X_{k|k-1}^{m_j}\}_{j=1}^J$ is a mini-batch of samples from the forecast ensemble $\{\rv X_{k|k-1}^m\}_{m=1}^M$, and $\hat{\bw}_{t}$ is a Monte Carlo approximation of Eq.~\eqref{eq:weight} such that

\s \begin{equation}\label{weight_app}
    \hat{\bw}_{t}\left(\bz, \bx_{k|k-1}^{m_j}\right) : = \frac{Q\left(\bz | \bz_{k|k-1}^{m_j} \right)}{\sum_{j=1}^J Q\left(\bz | \bx_{k|k-1}^{m'_j} \right)}.
\end{equation}

\s \noindent Having the estimated prior score $\hat{\bs}_{k|k-1}$, we can use Eq.~\eqref{S-posterior} to write the approximate posterior score as

\s \begin{equation}\label{Approx:S-posterior}
\hat{\bs}_{k | k}(\bz, t) := \hat{\bs}_{k|k-1}(\bz, t) + h(t) \nabla_\bx \log P(\by_k | \bz).
\end{equation}

\s \noindent With these approximations in mind, we are finally in a position to summarize the EnSF algorithm:

\newpage
\noindent\makebox[\linewidth]{\rule{\textwidth}{0.5pt}}\\
\vspace{-0.5cm}
\newline {\bf Algorithm}: \texttt{Ensemble Score Filter (EnSF)}\vspace{-0.2cm} \\ 
\noindent\makebox[\linewidth]{\rule{\textwidth}{0.5pt}}
\vspace{-0.3cm}
\newline1:\, {\bf Input}: The forecast model $\bvf$ and initial state pdf $P(\bx_0)$;
\vspace{0.1cm}
\newline2: Generate $M$ samples $\{\rv X_{0|0}^m\}_{m=1}^M$ from the initial pdf $P(\bx_0)$;
\vspace{0.1cm}
\newline3: \,{\bf for} $k = 1, 2, \ldots, K$: \quad\texttt{\% physical time loop}
\vspace{0.1cm}
\newline4: \qquad Run the forecast model $\bvf$ to get the forecast ensemble $\{\rv X_{k|k-1}^m\}_{m=1}^M$;
\vspace{0.1cm}
\newline5: \qquad {\bf for} $n = N-1, \ldots, 0$: \quad\texttt{\% pseudo-time loop for the backward SDE} 
\vspace{0.1cm}
\newline6: \qquad \quad\; Compute the weight $ \hat{\bw}_{t_{n+1}}(\bz_{t_{n+1}}^m, \bx_{k|k-1}^m)$ using Eq.~\eqref{weight_app};
\vspace{0.1cm}
\newline7: \qquad \quad\; Compute $\{\hat{\bs}_{k|k-1}(\bz_{t_{n+1}}^m, t_{n+1})\}_{m=1}^M$ using Eq.~\eqref{Approx:S-prior};
\vspace{0.1cm}
\newline8: \qquad \quad\; Compute and store $\{\hat{\bs}_{k|k}(\bz_{t_{n+1}}^m, t_{n+1})\}_{m=1}^M$ using Eq.~\eqref{Approx:S-posterior};
\vspace{0.1cm}
\vspace{0.1cm}
\newline9: \qquad\; {\bf end}\vspace{0.1cm} 
\newline10: \qquad\ Compute $\{\rv Z_{t_0}^m \}_{m=1}^M$ using Eq.~\eqref{Scheme:reverse} and set $\{\rv X_{k|k}^m\}_{m=1}^M \leftarrow \{\rv Z_{t_0}^m\}_{m=1}^M $; 
\vspace{0.1cm}
\newline11: {\bf end}\vspace{-0.1cm} \\
\noindent\makebox[\linewidth]{\rule{\textwidth}{0.5pt}}

%-=-=-=-=-=-=-=-=-=-=-=-=-=-=-=-=-=-=-=-=-=-=
%          EXPERIMENTAL DESIGN
%-=-=-=-=-=-=-=-=-=-=-=-=-=-=-=-=-=-=-=-=-=-=

\section{Experimental design} \label{sec:experimental_design}

Before presenting our results, we outline some details on the numerical model and reference EnKF method used in this study.

\subsection{Surface quasi-geostrophic (SQG) model}

The SQG model belongs to a special class of quasi-geostrophic models in which a fluid of constant potential vorticity (PV) is bounded between two flat, rigid surfaces \citep{tulloch_smith_2009a}. Despite its idealized nature, the system is capable of simulating turbulent motions similar to those occurring in real geophysical flows \citep{smith_et_al_2023}. It is also suitable for DA studies because the SQG flow is inherently chaotic and sensitive to initial condition errors \citep{rotunno_snyder_2008,durran_gingrich_2014}.

In this study, we adopt the SQG formulation proposed by \citet{tulloch_smith_2009b} in which the dynamics reduce to the nonlinear Eady model with an f-plane approximation as well as uniform stratification and shear. For this case, the governing equations simplify to the advection of potential temperature on the bounding surfaces. Those equations are solved numerically by first applying a fast Fourier transform (FFT) to map model variables to spectral space. They are integrated forward with a 4$^{\text{th}}$-order Runge Kutta scheme that uses a $2/3$ dealiasing rule and implicit treatment of hyperdiffusion. Further numerical details and open-source code can be found on the GitHub page shared at the end of the paper.

\subsection{Observing system simulation experiments}
The EnSF performance is assessed with $20$ ensemble members using a standard observing system simulation experiment (OSSE) framework where synthetic observations are generated by corrupting the true (nature) run with random noise. The construction of the nature run mostly follows the details of \citet{wang_et_al_2021}. Two notable exceptions are that we perform additional experiments with a higher resolution version of the model (96 $\times$ 96 points) and further carry out imperfect model experiments where the nature run is contaminated with unpredictable model errors.

More significant differences appear in the generation of synthetic observations. First, we utilize a 12-hour assimilation window. Compared to the 3-hour window of \citet{wang_et_al_2021}, this constitutes a more challenging scenario as it creates larger differences with the true state and likely causes more significant departures from Gaussianity in the forecast ensemble. Moreover, we test a wider range of observing networks, starting with the simpler case of a fully observed state with linear observations and finishing with a partially and nonlinearly observed state (50$\%$ coverage) that is additionally subject to unexpected model errors. More specific details about the different observing networks are deferred to Section \ref{sec:results}. 

\subsection{Reference LETKF method} \label{subsec:letkf}
While there is a vast array of available EnKF methods we could compare EnSF's performance against, here we opt for the Local Ensemble Transform Kalman Filter \citep[LETKF; ][]{hunt_et_al_2007}. LETKF is a popular EnKF variant that is currently implemented in several major operational centers \citep[e.g., ][]{schraff_et_al_2016,frolov_et_al_2024}. One of the most appealing features of this method is its efficiency -- the model state is decomposed into overlapping subdomains which are updated separately (and in parallel) using the analysis equations of the Ensemble Transform Kalman Filter \citep[ETKF; ][]{bishop_et_al_2001}. Our numerical implementation considers local regions surrounding single grid points defined by a cutoff radius. The cut-off radius is determined by standard Gaspari-Cohn (GC) taper functions \citep{gaspari_cohn_1999}, and we also apply the original $R$-localization strategy of \citet{hunt_et_al_2007}\footnote{With R-localization, the local observation error variances are multiplied by the inverse GC function.} to smoothly decrease the impact of observation away from each model grid point. Analogous to \citet{wang_et_al_2021}, the vertical and horizontal localization scales are coupled dynamically through the Rossby radius of deformation. We also apply the relaxation to prior spread (RTPS) inflation discussed in \citet{whitaker_hamill_2012} in order to prevent underdispersive analysis ensembles as a result of the finite ensemble size. In most experiments, the inflation factor is optimally tuned for LETKF while EnSF currently sets RTPS to 1.

While the simultaneous use of localization and inflation can greatly improve LETKF's performance, achieving optimal analysis results necessitates careful parameter tuning, which is computationally demanding in operational contexts. Moreover, sensitivity tests are required each time one makes changes to the NWP system (model resolution, type of assimilated observations, etc). Our initial results in Section \ref{sec:results} emphasize how LETKF's skill is highly sensitive to suboptimal choices of localization and inflation even in the conceivably simpler case of a fully and linearly observed state. This will be contrasted with EnSF's performance which leads to stable performance across all experiments and without any further ensemble regularization strategies (although additional experiments not presented in the paper suggest that optimal tuning of RTPS can further improve EnSF's performance). 

%-=-=-=-=-=-=-=-=-=-=-=-=-=-=-=-=-=-=-=-=-=-=
%               RESULTS
%-=-=-=-=-=-=-=-=-=-=-=-=-=-=-=-=-=-=-=-=-=-=
\section{Results} \label{sec:results}

The analysis skill of EnSF and LETKF is examined in the context of 4 different experimental settings with increasing complexity. Table \ref{Outline_Experiments} separates them according to the type of observations (linear vs.~nonlinear) as this accounts for the largest filtering differences.

\begin{table}[h!]
\centering
\begin{tabular}{>{\raggedright\arraybackslash}m{0.45\linewidth}|>{\raggedright\arraybackslash}m{0.45\linewidth}}
    \toprule
    \centering \textbf{Linear observations} & \hspace*{1.7cm} \textbf{Nonlinear observations} \\
    \midrule
    EXP\_L1: fully observed state & EXP\_NL1: fully observed state \\
    \addlinespace
    \makecell[lt]{EXP\_L2: fully observed state \\ subject to unknown model errors} & \makecell[lt]{EXP\_NL2: partially observed state \\ subject to unknown model errors} \\
    \bottomrule
\end{tabular}
\s \sO \caption{A summary of the 4 main experimental settings used to compare EnSF and LETKF in Section \ref{sec:results}. Note that EXP\_L1 contains more than one numerical test (e.g., using the SQG model at different resolutions).}
\label{Outline_Experiments}
\end{table}

\subsection{Linear observations}

In EXP\_L1, we compare the performance of EnSF and LETKF in the case where the state of the SQG model is fully and linearly observed. Measurements are corrupted by additive Gaussian errors such that

\vspace*{-3mm}
\begin{equation} \label{eq:fully_observed_state}
\rv Y_k = \rv X_k + \rv E_k^o\quad \text{with } \rv E_k^o \sim \mathcal{N}(\bo,\bR_k),
\end{equation}

\s \noindent with $\bR_k = \bI$ (i.e., the observation error variance is $1$K). While this represents a fairly straightforward test for most conventional DA methods, the combination of a small ensemble size ($20$) and large number of independent observations will likely cause weight collapse in the traditional (bootstrap) particle filter. This compels us to first explore the stability properties of the new EnSF method.  

\begin{figure}[h!]
\begin{center}
\includegraphics[scale = 0.55]{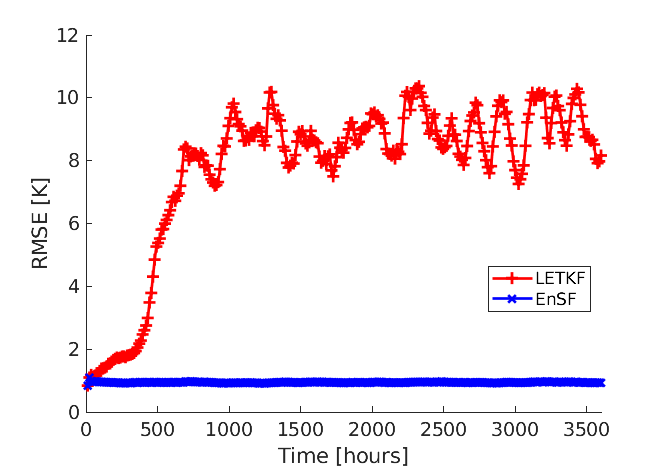}
\end{center}
\caption{Comparison of the analysis root mean squared errors (RMSEs) between EnSF and LETKF with a high-resolution configuration of the SQG model that uses $96 \times 96$ grid points in the horizontal. In this simulation, the SQG state is fully observed through linear observations with Gaussian errors (EXP\_L1). The LETKF experiment uses the same localization and inflation parameters as Figure 11a of \citet{wang_et_al_2021} [$L_h=2500$ km and RTPS $= 0.6$], while EnSF does not implement any localization and uses RTPS $= 1$.}
\label{Linear_96}
\end{figure}

It was already discussed in Section \ref{sec:experimental_design}\ref{subsec:letkf} that one of the major obstacles with EnKFs is the need for a careful parameter tuning. The impact of suboptimal choices for these parameters is illustrated in Figure ~\ref{Linear_96} where LETKF uses the same localization and inflation parameters as Fig.~11a of \citet{wang_et_al_2021} [horizontal localization scale of $L_h = 2500$ km and RTPS $=0.6$]. Despite the physical realism of these parameter choices, we see that the LETKF experiment quickly diverges due to changes in the model resolution ($96 \times 96$ vs.~$64 \times 64$ grid points in the horizontal), assimilation period (12h vs.~3h) and number of observations (fully vs.~partially observed state). By contrast, EnSF demonstrates stable performance throughout all 300 assimilation cycles without requiring any additional parameter tuning. 

\begin{figure}[ht!] %\vspace{-0.75em}
\begin{center}
\captionsetup[subfloat]{labelfont={large,bf},textfont=normalsize}
\subfloat[\Large{}]{\includegraphics[scale = 0.28]{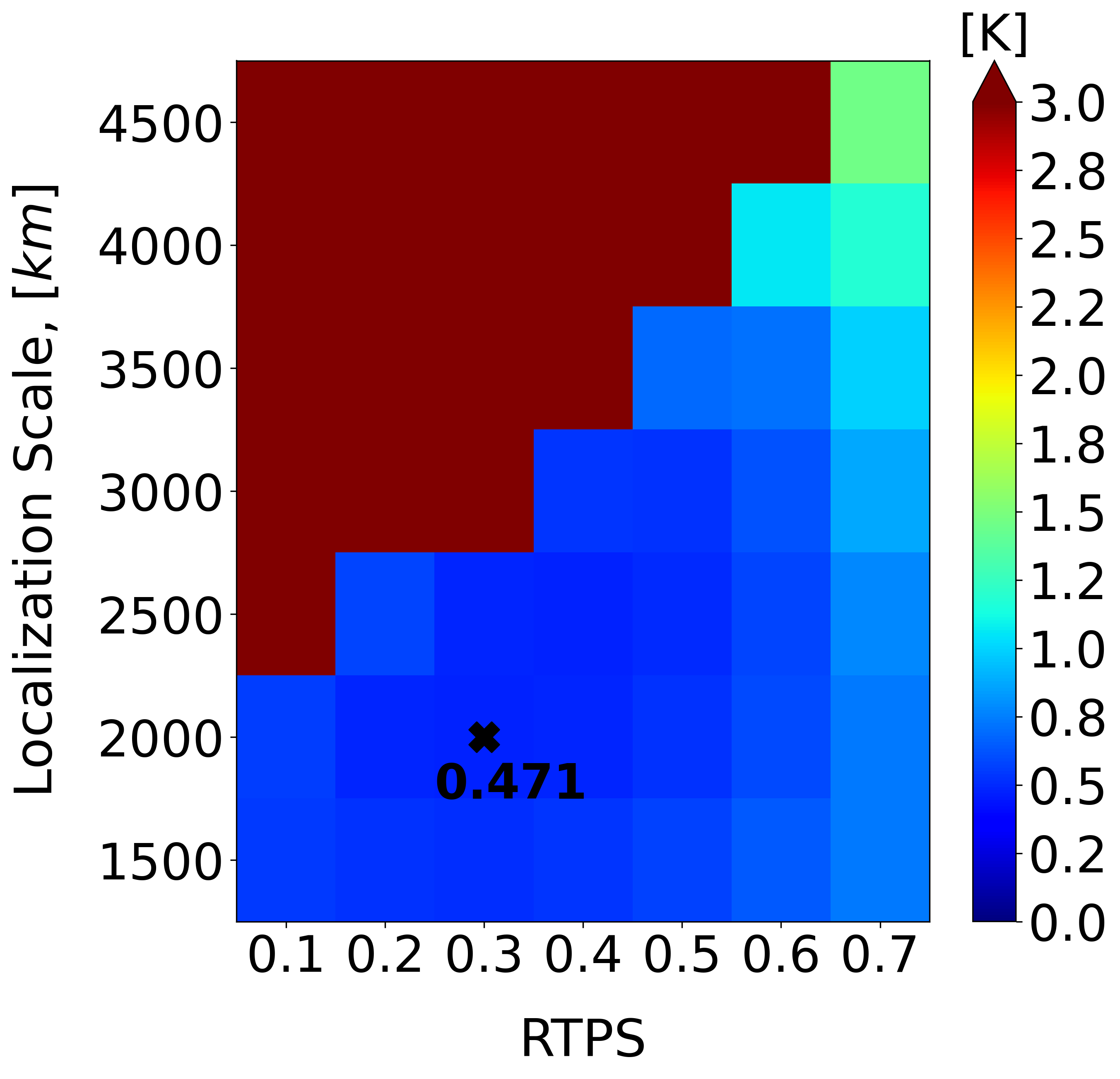} } \quad
\subfloat[]{\includegraphics[scale = 0.51]{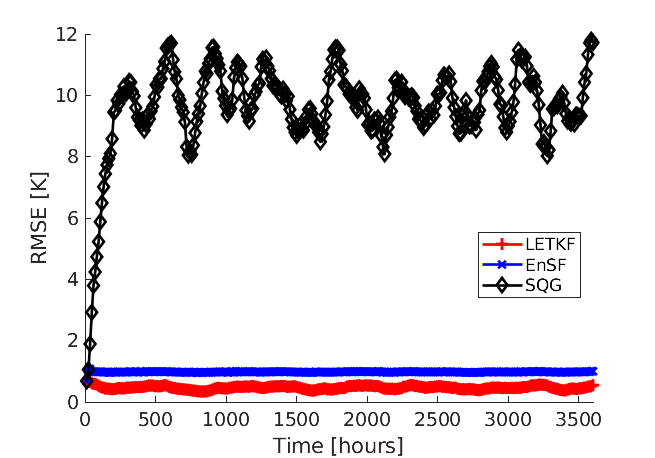} }
\end{center}
\caption{Same experimental setting as in Figure \ref{Linear_96} (EXP\_L1), but with the difference that the SQG model has a coarser grid made of $64 \times 64$ grid points in the horizontal. Panel (a) shows the time-averaged analysis RMSE errors of LETKF as a function of localization scale ($L_h$) and inflation (RTPS), with the best tuned experiment marked with a black cross. In addition to the analysis RMSE time series of EnSF and LETKF, the black curve in panel (b) shows the RMSE values of a reference experiment (SQG) in which no observations are assimilated.}
\label{Linear_64}
\end{figure}

To achieve a fair comparison between the two DA methods in this linear observation regime, our next step is to optimize LETKF's localization and inflation parameters. We revert back to the coarser model resolution of $64 \times 64$ horizontal grid points and see from Figure \ref{Linear_64}a that the best LETKF performance is achieved with $L_h = 2000$ km and RTPS = $0.3$. Evidently, LETKF performs slightly better in this case. However, it is worth noting that EnSF continues to maintain stable performance despite the change in model resolution and lack of additional parameter tuning. EnSF's sensitivity to different parameter settings will be the subject of future work and is expected to further reduce the gap between the two filtering techniques in this linear regime (EXP\_L1).

\begin{figure}[ht!] %\vspace{-0.75em}
\begin{center}
\captionsetup[subfloat]{labelfont={large,bf},textfont=normalsize}
\subfloat[Truth]{\includegraphics[scale = 0.51]{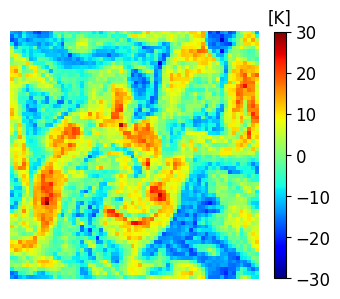} } \qquad \quad
\subfloat[$20\%$ noise]{\includegraphics[scale = 0.51]{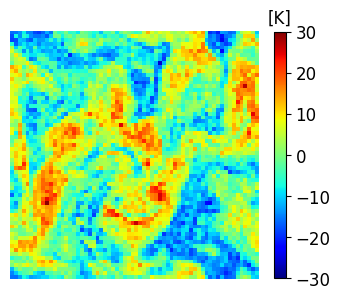} } \\
\subfloat[$30\%$ noise]{\includegraphics[scale = 0.51]{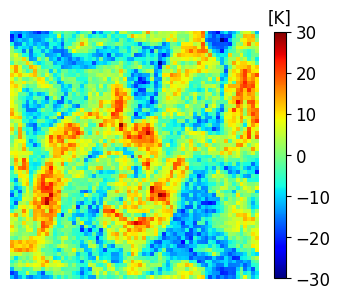} } \qquad \quad
\subfloat[$40\%$ noise]{\includegraphics[scale = 0.51]{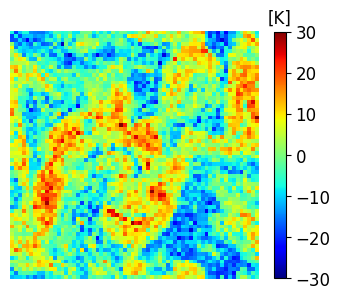} } \qquad \quad
\subfloat[$50\%$ noise]{\includegraphics[scale = 0.51]{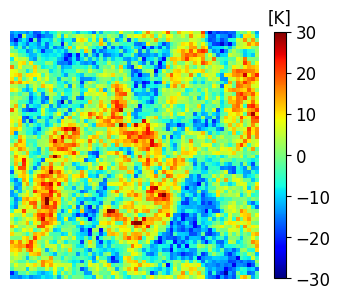} }
\end{center}
\caption{Visualizing the impacts from corrupting the true SQG state with model errors of different amplitude (relevant to EXP\_L2 and EXP\_NL2). Please refer to the main text for more details on how these model errors are incorporated in our DA experiments.}
\label{fig:demo_noise}
\end{figure}

In the next experimental setting EXP\_L2, we address the more challenging but realistic scenario involving an imperfect model. While still assuming direct measurements of all state variables, we add an a-priori \textit{unknown} stochastic noise $\rv E^{u}$ to the SQG state evolution. Assuming the noise is Gaussian, the modified stochastic-dynamic model can be written as
\begin{equation} \label{eq:imperfect_model}
\rv X_k = \bvf_{k-1}(\rv X_{k-1}, \rv E^m_{k-1}) + \rv E^{u}_{k}\quad \text{with } \rv E^u_k \sim \mathcal{N}(\bo,\bE_k).
\end{equation}
The covariance matrix $\bE_k$ is taken to be diagonal as we want to impose spatially uncorrelated model noise. We also assume that the unknown model errors are white in time (i.e., no temporal correlations). More specifically, $\rv E^{u}_{k}$ in EXP\_L2 is defined as the composition of four distinct state-dependent error processes: (1) $20\%$ chance of occurrence (in time) with $20\%$ model noise, (2) $15\%$ chance of occurrence with $30\%$ model noise, (3) $10\%$ chance of occurrence with $40\%$ model noise, and (4) $5\%$ chance of occurrence with $50\%$ model noise. The purpose of introducing this variety of model errors is to simulate the effects of flow-dependent model uncertainties. For example, the rare occurrence of high-amplitude model errors in the 4$^{\text{th}}$ process might be associated with the rapid development of dynamical instabilities in the $\theta$ field. The study of \citet{held_et_al_1995} shows examples of such instabilities and discusses how they tend to form along temperature filaments (see their Fig.~2).

It is crucial to highlight that although we specified a particular structure for the time-dependent error process $\rv E^{u}_{k}$, obtaining an explicit expression for $\rv E^{u}_{k}$ is impractical in reality. Consequently, adjusting the LETKF inflation and localization parameters to account for the unknown model errors $\rv E^u_{k}$ is not feasible. In Figure \ref{fig:demo_noise}, we visually illustrate the impact of adding model errors of various amplitude to the true state at the final time.

\begin{figure}[h!]
\begin{center}
\includegraphics[scale = 0.55]{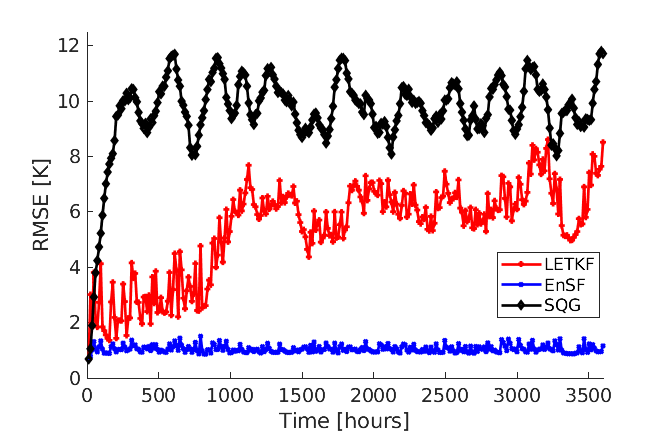}
\end{center}
\caption{Same as Figure \ref{Linear_64} but with the exception that the presented comparison refers to the imperfect model setting EXP\_L2. Note that the LETKF experiment (red curve) uses the optimal localization and inflation values from Figure \ref{Linear_64}a.}
\label{Linear_shocks_SQG}
\end{figure}

\begin{figure}[h!]
\begin{center}
\captionsetup[subfloat]{labelfont={large,bf},textfont=normalsize}
\subfloat[Truth]{\includegraphics[scale = 0.52]{SQG_nature_run.png} } \qquad \quad
\subfloat[EnSF analysis]{\includegraphics[scale = 0.52]{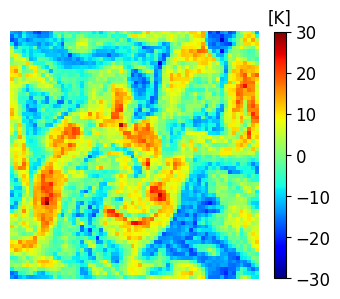} } \qquad \quad
\subfloat[LETKF analysis]{\includegraphics[scale = 0.52]{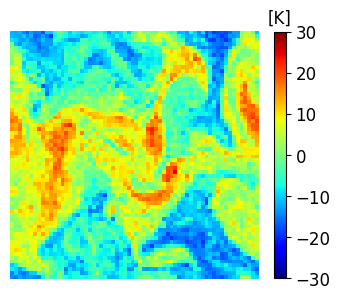} } \\
\subfloat[EnSF error]{\includegraphics[scale = 0.53]{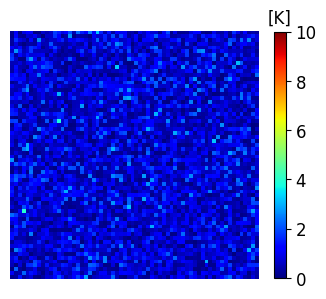} }\qquad \quad
\subfloat[LETKF error]{\includegraphics[scale = 0.53]{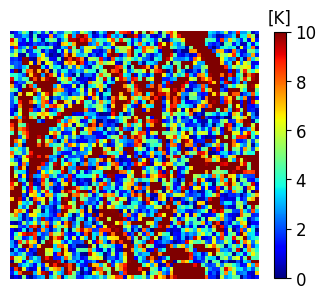} }
\end{center}
\caption{The top row shows the analysis ensemble means from EnSF and LETKF under scenario EXP\_L2 and with respect to the true potential temperature field at the final observation time. The analysis mean errors of the two experiments are displayed on the bottom row.}
\label{Linear_2D_Error}
\end{figure}

A comparison between the analysis RMSE errors of EnSF and LETKF in EXP\_L2 is presented in Figure \ref{Linear_shocks_SQG}. Unlike our previous experiment, LETKF diverges from the true state as model errors accumulate in time, with analysis RMSEs now comparable to the free forecast run where no observations are assimilated. This finding offers additional evidence for the enhanced sensitivity of LETKF to model imperfections -- without the implementation of more advanced regularization techniques (e.g., adaptive inflation), LETKF will be susceptible to analysis errors even when nearly optimal parameters have been used and the system is fully and linearly observed. This is to be contrasted with EnSF's performance where we achieve stable performance throughout all analysis cycles despite the lack of any further tuning. A snapshot of these differences at the last analysis time is shown in Figure \ref{Linear_2D_Error}. The top 3 figures offer a visual confirmation that EnSF's analysis mean (panel b) is much closer to the truth (panel a). While LETKF (panel c) does well at representing large-scale patterns, it struggles to capture some of the small-scale features and extreme values in the potential temperature field. The larger error magnitude in LETKF is also confirmed by the two error plots in the second row.

\subsection{Nonlinear observations}

In the second round of experiments, the synthetically generated observations are nonlinear functions of the SQG state. Specifically, we consider an arctangent nonlinearity and additive Gaussian observation errors. The observation model relevant to EXP\_NL1 from Table \ref{Outline_Experiments} can be written as

$$\rv Y_{k} = \arctan(\rv X_{k}) + \tilde{\rv E}^o_{k}\quad \text{with } \tilde{\rv E}_k^o \sim \mathcal{N}(\bo, \tilde{\bR}_k).$$
Notice that the observation error variance is scaled relative to the linear observation experiments in order to reflect the narrower range of the arctangent function. In this case, we choose $\tilde{\bR}_k = 0.01 \times \bR_k = 0.01 \times \bI$. Moreover, EXP\_NL1 tests are carried out in the absence of unpredictable model shocks.

\begin{figure}[h!]
\begin{center}
\captionsetup[subfloat]{labelfont={large,bf},textfont=normalsize}
\subfloat[]{\includegraphics[scale = 0.26]{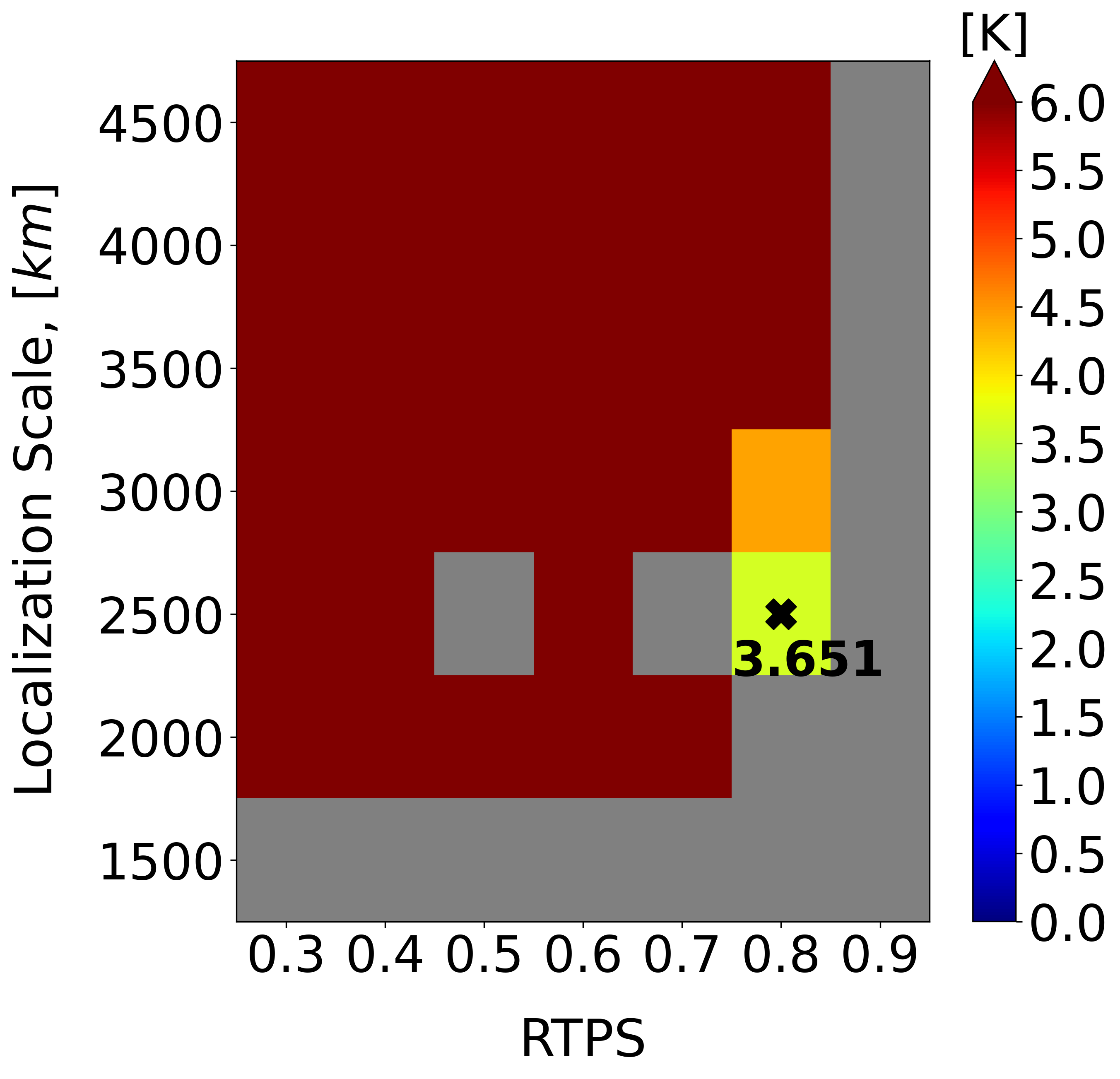} } \qquad
\subfloat[]{\includegraphics[scale = 0.53]{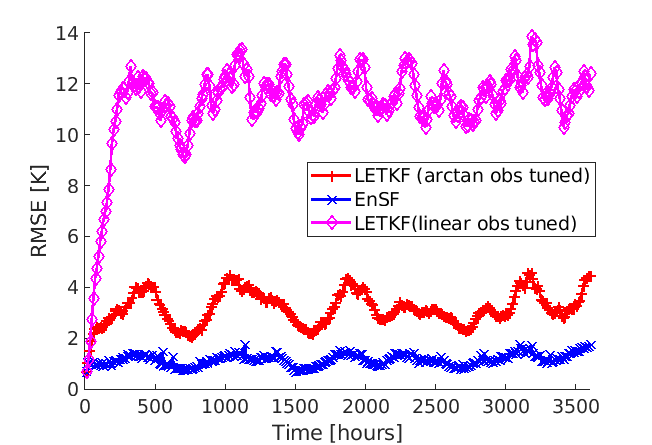}}
\end{center}
\caption{Filtering performance in the case of a fully observed state with an arctangent nonlinearity (EXP\_NL1). The format of this Figure is close to Figure \ref{Linear_64} with the difference that the reference experiment in magenta line is LETKF with localization and inflation parameters optimally tuned for a linear observing system.}
\label{Arctan_Full}
\end{figure}

Analogous to the linear observation setting, stabilizing LETKF's performance requires extensive parameter tuning. Figure \ref{Arctan_Full}a indicates the existence of a very narrow range of optimal localization and inflation parameters for which LETKF does not diverge -- a manifestation of the strong deviations from non-Gaussianity in the posterior distribution. The red curve in Figure \ref{Arctan_Full}b confirms that this best tuned LETKF maintains stability during the entire experiment. On the other hand, a naive substitution of the optimal LETKF parameters from the linear observation setting results in a rapidly diverging filter (magenta curve in Figure \ref{Arctan_Full}b). While this comparison admittedly constitutes an extreme modification of the observing system, it reaffirms our earlier point that any changes in operational NWP systems require costly calibrations of the underlying EnKF algorithm. By contrast, EnSF systematically outperforms the best LETKF configuration without further application of ensemble regularization strategies. 

\begin{figure}[h!]
\begin{center}
\captionsetup[subfloat]{labelfont={large,bf},textfont=normalsize}
\subfloat[]{\includegraphics[scale = 0.73]{SQG_nature_run.png} } \quad\quad
\subfloat[]{\includegraphics[scale = 0.73]{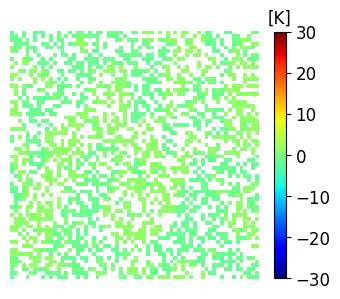} }
\end{center}
%\vspace{-0.75em}
\caption{(a) The original SQG state prior to the addition of model errors. (b) A snapshot from a partial observing network with a 50\% coverage showing the loss of information (\textit{cf.}~limited range of values) due to the application of an arctangent nonlinearity.}
\label{Arctan_observations}
\end{figure}

Our last experimental setting EXP\_NL2 considers the most complex DA scenario -- only half of the SQG state is observed nonlinearly such that

$$\rv Y_{k} = \bH_k \arctan(\rv X_{k}) + \tilde{\rv E}^o_{k}\quad \text{with } \tilde{\rv E}_k^o \sim \mathcal{N}(\bo, \tilde{\bR}_k),$$

\noindent where $\bH_k$ is a selection matrix determining which state variables are observed at a given time level $k$. Similar to \citet{wang_et_al_2021}, we implement a procedure which randomly selects the observed model grid points for each $k$. An example realization of the nonlinear observations is presented in Figure \ref{Arctan_observations}b. Comparison with the true field in Figure \ref{Arctan_observations}a serves as a good illustration for the squashing effect of the arctangent nonlinearity, and the general difficulty of extracting useful information from observations with a severely limited numerical range.

We also work under the imperfect model assumption described by Eq.~\eqref{eq:imperfect_model}, but choose a simpler definition for the unknown model errors $\rv E^{u}_k$. In particular, a single stochastic noise is added 10\% of the time with a magnitude that equals 30\% of the nature run's values.

\begin{figure}[h!]
\begin{center}
\captionsetup[subfloat]{labelfont={large,bf},textfont=normalsize}
\subfloat[]{\includegraphics[scale = 0.28]{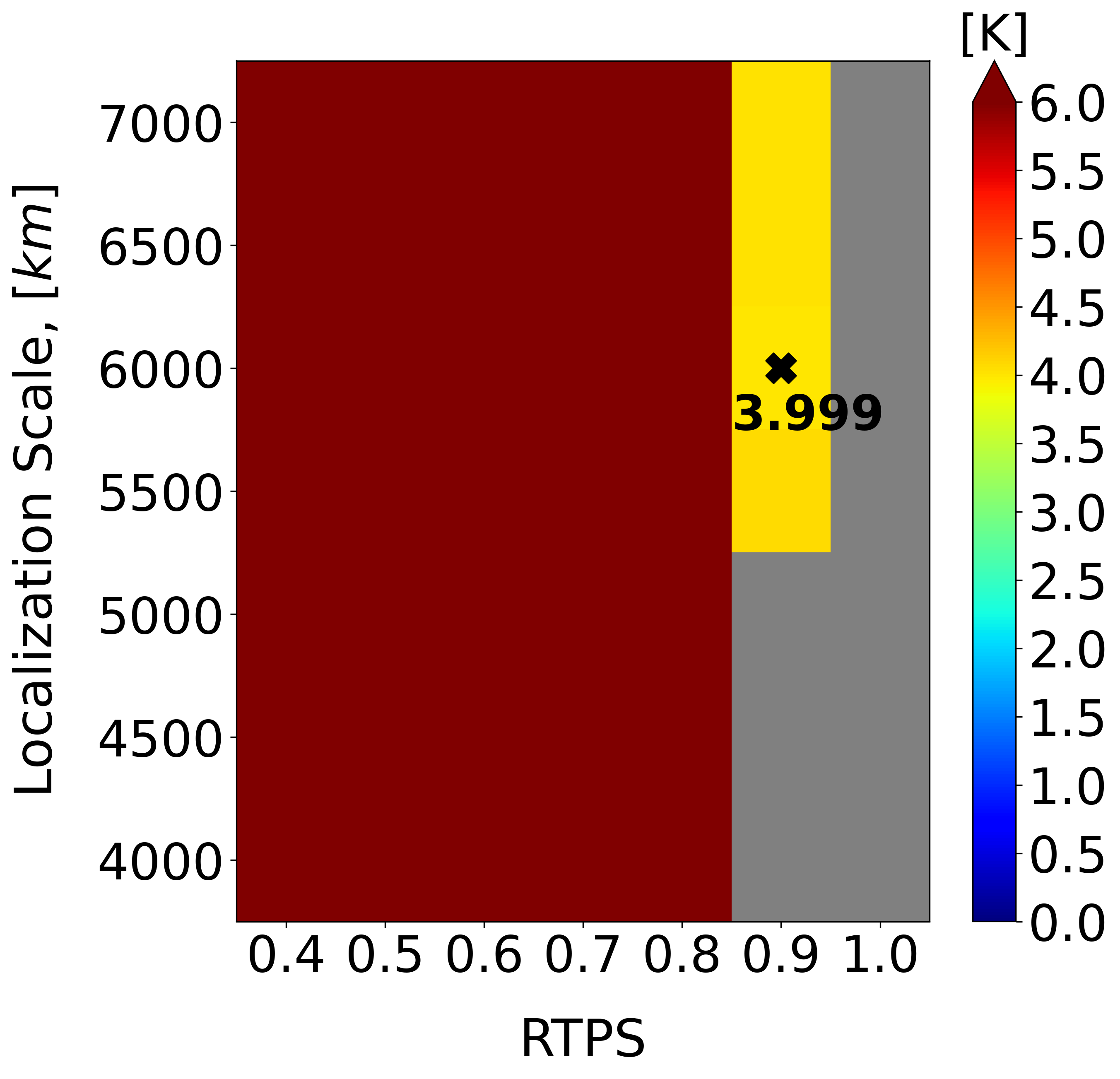} } \quad
\subfloat[]{\includegraphics[scale = 0.54]{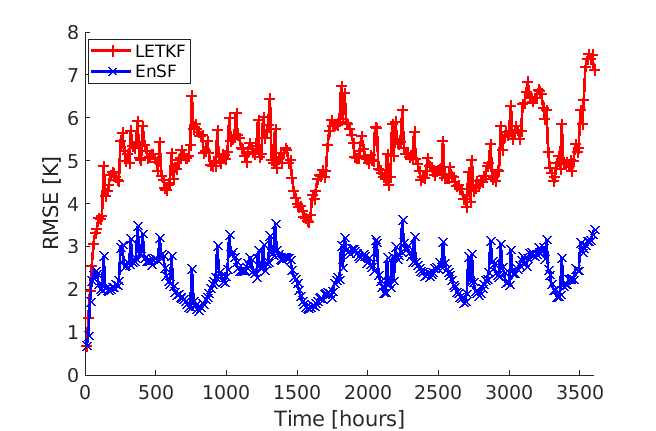} }\end{center}
\caption{Same as Figure \ref{Arctan_Full} but the analysis RMSEs refer to EXP\_NL2, which contains a partially observed state that is additionally subject to imperfect model assumptions.}
\label{Arctan_50_obs}
\end{figure}

Under the challenging DA setting of EXP\_NL2, we find that EnSF's analysis RMSEs remain nearly intact (compare blue curves in Figures \ref{Arctan_Full}b and \ref{Arctan_50_obs}b), while LETKF (Figure \ref{Arctan_50_obs}a) exhibits a visible skill deterioration (compare red curves in Figures \ref{Arctan_Full}b and \ref{Arctan_50_obs}b) despite our efforts to fine tune its performance for this case.

\begin{figure}[!ht]
\begin{center}
\captionsetup[subfloat]{labelfont={large,bf},textfont=normalsize}
\subfloat[]{\includegraphics[scale = 0.26]{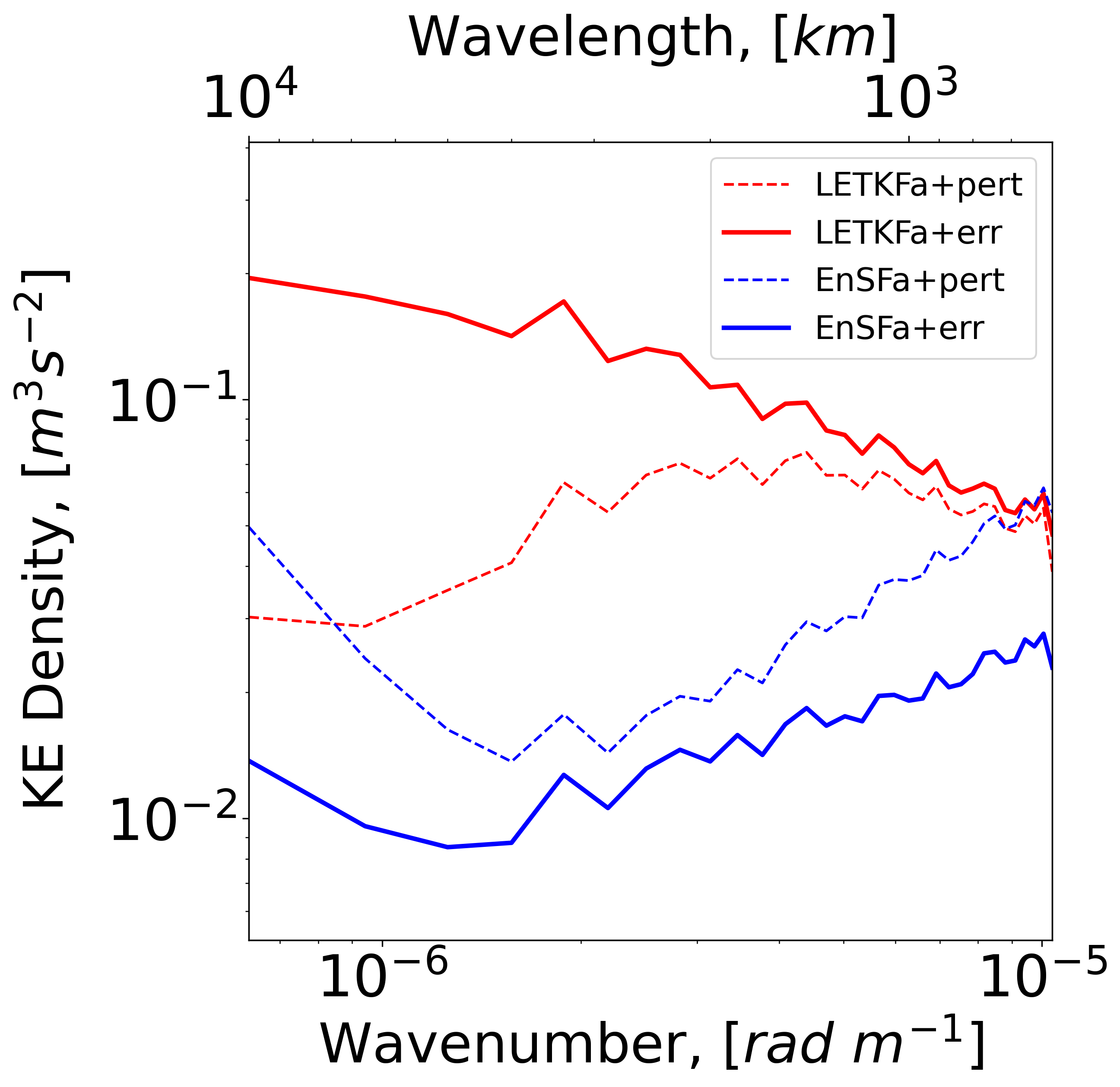} } 
\subfloat[]{\includegraphics[scale = 0.26]{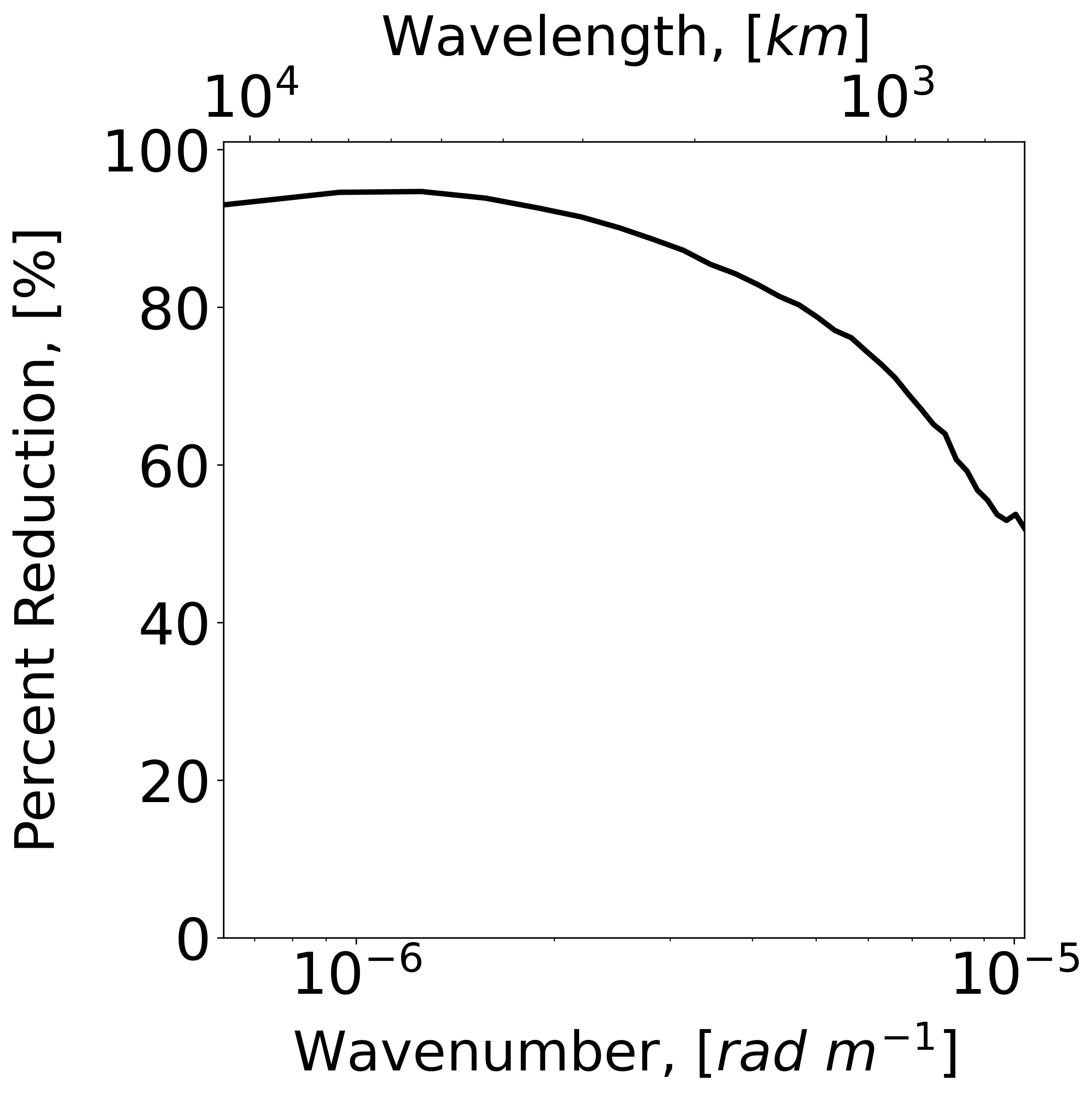} } 
\subfloat[]{\includegraphics[scale = 0.26]{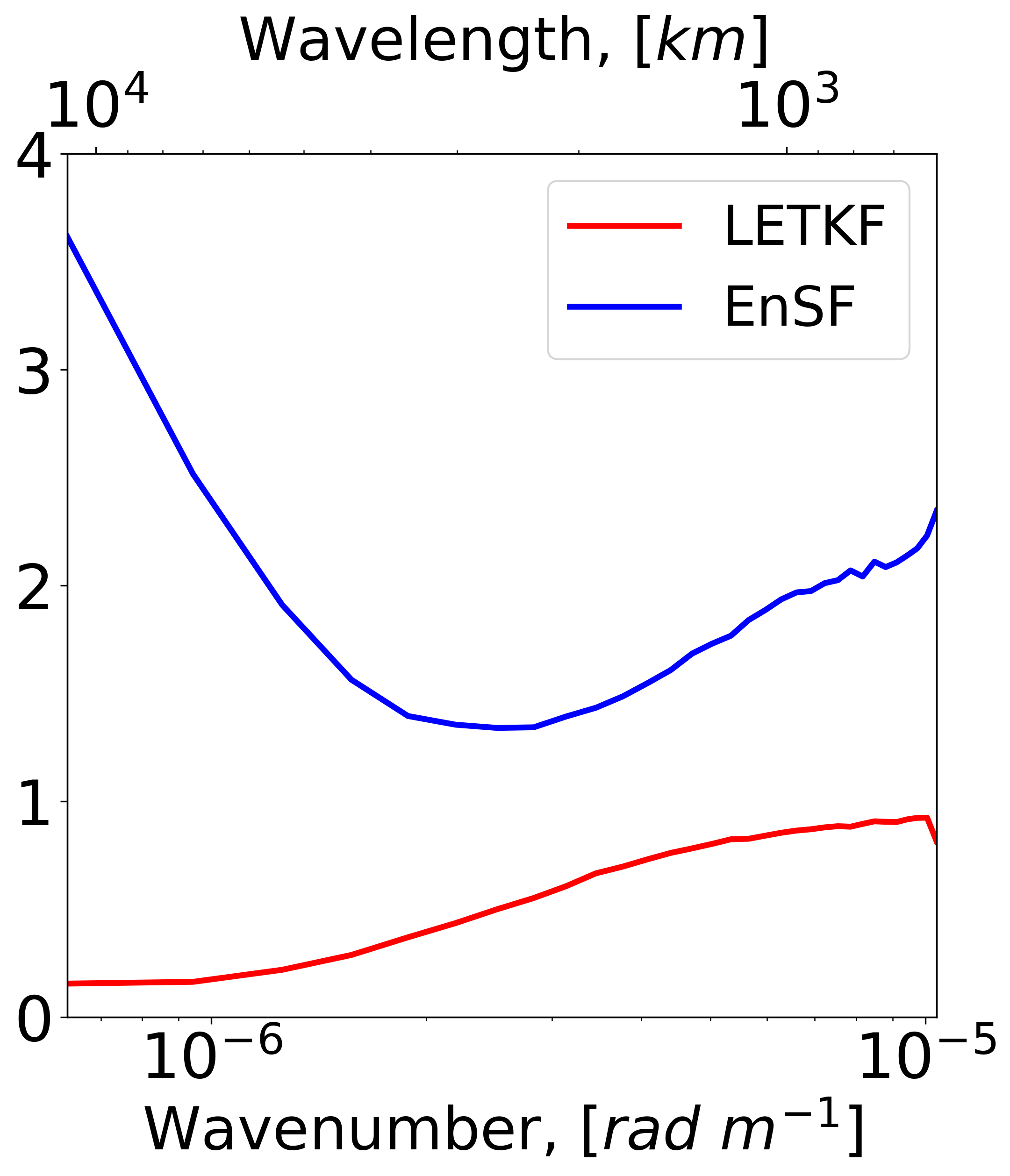}}
\end{center}
  \caption{Spectral diagnostics associated with EXP\_NL2. (a) Time-averaged kinetic energy (KE) density of the analysis mean errors (solid curves) and analysis perturbations (dashed curves) for EnSF (blue) and LETKF (red). (b) Percent improvement of EnSF's analysis mean errors as a function of wavenumber. (c) Analysis consistency ratios for the 2 ensemble filters. Note that the values are obtained by dividing the dashed curves by the solid curves in panel (a).}
\label{fig:spectral_diags}
\end{figure}

Given the multiscale characteristics of the SQG model, we conclude this section by presenting a spectral comparison of the EnSF and LETKF results in Figure \ref{fig:spectral_diags}. The plotted diagnostics still refer to the EXP\_NL2 setting (partially and nonlinearly observed state corrupted by unpredictable model errors) but are now averaged in time. Examination of the solid red and blue curves in panel (a) indicates that EnSF has consistently lower analysis errors for all wavenumbers, but the most significant improvements occur at large scales (also refer to panel b). Focusing our attention on the LETKF results (red cuferves), we notice a systematic underdispersion of the analysis ensemble, which is most pronounced at large scales (panel c). This is perhaps one of the important reasons why a large number of LETKF experiments diverge in EXP\_NL2. A different situation emerges for EnSF (blue curves) which tends to be overdispersive on average (panel c). This behavior is connected to the specific inflation settings used in EnSF (RTPS $=1$) but is likely also influenced by our choice of SDE and score parameters such as $b(t), \sigma (t)$ and $h(t)$. We hypothesize that an optimal selection of these parameters will further improve EnSF's performance and lead to a better spread-error consistency. Nevertheless, it is clear that even this ``vanilla'' implementation of EnSF offers significant performance benefits over LETKF. 

%-=-=-=-=-=-=-=-=-=-=-=-=-=-=-=-=-=-=
%           CONCLUSIONS
%-=-=-=-=-=-=-=-=-=-=-=-=-=-=-=-=-=-=
\section{Conclusions} \label{sec:conclusions}
In this study, we introduced a stable and highly efficient implementation of the Ensemble Score Filter (EnSF) for sequential data assimilation with geophysical systems. The theoretical basis for EnSF comes from diffusion models -- an extremely popular class of generative AI (GenAI) methods which have the ability to produce highly realistic images and videos. Like other diffusion-based techniques, EnSF uses score functions (the gradient of the log probability) to store complex information about the underlying filtering (posterior) distribution. However, sampling from the posterior utilizes a training-free, Monte Carlo procedure which only requires access to the forecast ensemble members.

The resulting algorithm is applied to the surface quasi-geostrophic (SQG) model and compared against a benchmark Local Ensemble Transform Kalman Filter (LETKF). The analysis performance of both methods is explored in a hierarchy of experiments with increasing complexity. While they demonstrate comparable skill in the case of a fully and linearly observed state, EnSF performs significantly better when observations are partial and nonlinear, as well as when the model is subject to unexpected errors. We report stable EnSF performance in all experiments despite the lack of localization and additional parameter tuning. Even though we only use 20 ensemble members in our experiments, the results suggest that EnSF can reliably capture the non-Gaussian characteristics of a high-dimensional state with $8,192$ components. Although LETKF is competitive or even slightly more skillful in the linear observation case, the results are highly sensitive to the choice of inflation and localization parameters; this effect is even more pronounced with nonlinear observations when slight changes in the LETKF settings often lead to rapid filter divergence.

The EnSF findings outlined in this study are very encouraging and suggest a few possible avenues for further development. An immediate extension of this work will be to explore the sensitivity of EnSF to different types of observing networks. Moreover, we believe the algorithm will benefit from additional theoretical refinements. One ongoing line of research from the authors is to derive an optimal damping function $h(t)$. Given the important role of $h(t)$ in determining how observations are incorporated in the diffusion-based filtering procedure (\textit{cf.}~Eq.~\ref{S-posterior}), we expect to see further improvements in EnSF's analysis performance. Although localization was not required to prevent filter divergence in the SQG context, more work is still needed to understand how sampling errors affect EnSF in higher dimensions. Together with a suitable choice for $h(t)$, an optimally designed localization scheme is expected to make EnSF comparable to a fine-tuned LETKF in linear-Gaussian regimes.

%-=-=-=-=-=-=-=-=-=-=-=-=-=-=-=-=-=-=
%            FINISH UP
%-=-=-=-=-=-=-=-=-=-=-=-=-=-=-=-=-=-=
\vspace*{-1mm}
\subsubsection{Acknowledgements.} This material is based upon work supported by the U.S. Department of Energy, Office of Science, Office of Advanced Scientific Computing Research, Applied Mathematics program under the contract ERKJ387 at the Oak Ridge National Laboratory, which is operated by UT-Battelle, LLC, for the U.S. Department of Energy under Contract DE-AC05-00OR22725. The first author (FB) would also like to acknowledge support from the U.S. National Science Foundation through project DMS-2142672 and the support from the U.S. Department of Energy, Office of Science, Office of Advanced Scientific Computing Research, Applied Mathematics program under Grant DE-SC0022297. The computing for this project was performed on the high performance computing (HPC) cluster at the Florida State University Research Computing Center.

\vspace*{-1mm}
\subsubsection{Data availability statement.}
All code written as part of this study will be made available on GitHub upon completing the peer-review process for this article. An open-source version of the SQG model used in our experiments can be found at \url{https://github.com/jswhit/sqgturb}.

%%% References
\newpage
\renewcommand\bibname{\Large{References}}
\bibliographystyle{ametsoc2014}
%\bibliography{references}

%%% Finalize
\end{document}

%% file: macros.tex
\usepackage[T1]{fontenc}
\usepackage{amsfonts,amsmath,amssymb}
\usepackage{mathtools}
\usepackage{mathrsfs}
\usepackage{upgreek}
\usepackage{cancel}
\usepackage{makecell} % long text in table cells
\usepackage{booktabs} % for prettier tables
\usepackage{makecell} % for line breaks in table cells
\usepackage{array}    % for table column formatting
\usepackage[hidelinks]{hyperref}
\usepackage[sectionbib]{natbib} 
\setcitestyle{aysep={}} % remove "," between authors and year with in-text citations
\usepackage{doi}
\usepackage{xcolor}
\usepackage{graphicx}
\usepackage[caption=false]{subfig}
\usepackage{algorithm}
\usepackage{algorithmic}
\usepackage{fancyhdr} % format document headers
\usepackage{enumitem}

\hypersetup{
  colorlinks=true,
  linkcolor=black,
  urlcolor=blue,
  citecolor=black,
  filecolor=black
}

\newcommand{\s}{{\vspace*{3mm}}} 
\newcommand{\sO}{{\vspace*{1mm}}} 
\newcommand{\sT}{{\vspace*{2mm}}} 
\newcommand{\sR}{{\mathbb{R}}}

\newcommand{\bx}{{\mathbf{x}}}
\newcommand{\by}{{\mathbf{y}}}
\newcommand{\bz}{{\mathbf{z}}}

\newcommand{\bw}{{\mathbf{w}}}

\newcommand{\bo}{{\mathbf{0}}}

\newcommand{\bs}{{\mathbf{s}}}
\newcommand{\bvf}{{\mathbf{f}}}
\newcommand{\bvh}{{\mathbf{h}}}
\newcommand{\T}{{^\top}}
\newcommand{\bR}{{\mathbf{R}}}

\newcommand{\bH}{{\mathbf{H}}}
\newcommand{\bI}{{\mathbf{I}}}

\newcommand{\bE}{{\mathbf{\Sigma}}}

\newcommand{\rv}[1]{\boldsymbol{#1}}

\usepackage{hyperref}
\usepackage{xcolor}
\hypersetup{
    colorlinks,
    linkcolor={black},
    linkcolor={black},
    citecolor={black},
    urlcolor={blue!80!black}
}

%% file: preprint.bbl
\begin{thebibliography}{48}
\providecommand{\natexlab}[1]{#1}
\providecommand{\url}[1]{\texttt{#1}}
\renewcommand{\UrlFont}{\rmfamily}
\providecommand{\urlprefix}{URL }
\expandafter\ifx\csname urlstyle\endcsname\relax
  \providecommand{\doi}[1]{doi:\discretionary{}{}{}#1}\else
  \providecommand{\doi}{doi:\discretionary{}{}{}\begingroup
  \urlstyle{rm}\Url}\fi
\providecommand{\eprint}[2][]{\url{#2}}

\bibitem[{Aksoy et~al.(2009)Aksoy, Dowell,, and Snyder}]{aksoy_et_al_2009}
Aksoy, A., D.~Dowell, and C.~Snyder, 2009: A multicase comparative assessment
  of the ensemble {Kalman} filter for assimilation of radar observations. {Part
  I: Storm-scale} analyses. \textit{Mon.\ Wea.\ Rev.}, \textbf{137},
  1805–1824, \doi{10.1175/2008MWR2691.1}.

\bibitem[{Aksoy et~al.(2010)Aksoy, Dowell,, and Snyder}]{aksoy_et_al_2010}
Aksoy, A., D.~Dowell, and C.~Snyder, 2010: A multicase comparative assessment
  of the ensemble {K}alman filter for assimilation of radar observations. {Part
  II}: {Short}-range ensemble forecasts. \textit{Mon.\ Wea.\ Rev.},
  \textbf{138}, 1273--1292, \doi{10.1175/2009MWR3086.1}.

\bibitem[{Anderson(2001)}]{anderson_2001}
Anderson, J.~L., 2001: An ensemble adjustment {K}alman filter for data
  assimilation. \textit{Mon.\ Wea.\ Rev.}, \textbf{129}, 2884--2903,
  \doi{10.1175/1520-0493(2001)129<2884:AEAKFF>2.0.CO;2}.

\bibitem[{Anderson(2022)}]{anderson_2022}
Anderson, J.~L., 2022: A quantile-conserving ensemble filter framework. {P}art
  {I}: {U}pdating an observed variable. \textit{Mon.\ Wea.\ Rev.},
  \textbf{150}, 1061–1074, \doi{10.1175/MWR-D-21-0229.1}.

\bibitem[{Anderson(2023)}]{anderson_2023}
Anderson, J.~L., 2023: A quantile-conserving ensemble filter framework. {Part
  II: R}egression of observation increments in a probit and probability
  integral transformed space. \textit{Mon.\ Wea.\ Rev.}, \textbf{151},
  2759–2777, \doi{10.1175/MWR-D-23-0065.1}.

\bibitem[{Bao et~al.(2016)Bao, Cao, Meir,, and Zhao}]{BDSDE}
Bao, F., Y.~Cao, A.~Meir, and W.~Zhao, 2016: A first order scheme for backward
  doubly stochastic differential equations. \textit{SIAM/ASA J. Uncertain.
  Quantif.}, \textbf{4~(1)}, 413--445.

\bibitem[{Bao et~al.(2023{\natexlab{a}})Bao, Zhang,, and Zhang}]{EnSF_2023}
Bao, F., Z.~Zhang, and G.~Zhang, 2023{\natexlab{a}}: An ensemble score filter
  for tracking high-dimensional nonlinear dynamical systems. \textit{arXiv},
  1--17, \doi{arXiv:2309.00983}.

\bibitem[{Bao et~al.(2023{\natexlab{b}})Bao, Zhang,, and Zhang}]{SF_2023}
Bao, F., Z.~Zhang, and G.~Zhang, 2023{\natexlab{b}}: A score-based nonlinear
  filter for data assimilation. \textit{arXiv}, 1--20, \doi{arXiv:2306.09282}.

\bibitem[{Bao et~al.(2023{\natexlab{c}})Bao, Zhang,, and Zhang}]{UF_2023}
Bao, F., Z.~Zhang, and G.~Zhang, 2023{\natexlab{c}}: A unified filter method
  for jointly estimating state and parameters of stochastic dynamical systems
  via the ensemble score filter. \textit{arXiv}, 1--24, \doi{arXiv:2312.10503}.

\bibitem[{Baranchuk et~al.(2022)Baranchuk, Voynov, Rubachev, Khrulkov,, and
  Babenko}]{baranchuk2022labelefficient}
Baranchuk, D., A.~Voynov, I.~Rubachev, V.~Khrulkov, and A.~Babenko, 2022:
  Label-efficient semantic segmentation with diffusion models.
  \textit{International Conference on Learning Representations},
  \urlprefix\url{https://openreview.net/forum?id=SlxSY2UZQT}.

\bibitem[{Bishop et~al.(2001)Bishop, Etherton,, and
  Majumdar}]{bishop_et_al_2001}
Bishop, C.~H., B.~J. Etherton, and S.~J. Majumdar, 2001: Adaptive sampling with
  the ensemble transform {K}alman filter. {Part I: Th}eoretical aspects.
  \textit{Mon.\ Wea.\ Rev.}, \textbf{129}, 420–436,
  \doi{10.1175/1520-0493(2001)129<0420:ASWTET>2.0.CO;2}.

\bibitem[{Chan et~al.(2020)Chan, Anderson,, and Chen}]{chan_et_al_2020}
Chan, M.-Y., J.~Anderson, and X.~Chen, 2020: An efficient bi-{G}aussian
  ensemble {K}alman filter for satellite infrared radiance data assimilation.
  \textit{Mon.\ Wea.\ Rev.}, \textbf{148}, 5087–5104,
  \doi{10.1175/MWR-D-20-0142.1}.

\bibitem[{Chipilski(2023)}]{chipilski_2023}
Chipilski, H.~G., 2023: Exact nonlinear state estimation. \textit{arXiv},
  1–31, \doi{10.48550/arXiv.2310.10976}.

\bibitem[{Chipilski et~al.(2020)Chipilski, Wang,, and
  Parsons}]{chipilski_et_al_2020}
Chipilski, H.~G., X.~Wang, and D.~B. Parsons, 2020: Impact of assimilating
  {PECAN} profilers on the prediction of bore-driven nocturnal convection: {A}
  multiscale forecast evaluation for the 6 july 2015 case study. \textit{Mon.\
  Wea.\ Rev.}, \textbf{148}, 1147–1175, \doi{10.1175/MWR-D-19-0171.1}.

\bibitem[{Chipilski et~al.(2022)Chipilski, Wang, Parsons, Johnson,, and
  Degelia}]{chipilski_et_al_2022}
Chipilski, H.~G., X.~Wang, D.~B. Parsons, A.~Johnson, and S.~K. Degelia, 2022:
  The value of assimilating different ground-based profiling networks on the
  forecasts of bore-generating nocturnal convection. \textit{Mon.\ Wea.\ Rev.},
  \textbf{150}, 1273–1292, \doi{10.1175/MWR-D-21-0193.1}.

\bibitem[{Crisan and Doucet(2002)Crisan, and Doucet}]{crisan_doucet_2002}
Crisan, D., and A.~Doucet, 2002: A survey of convergence results on particle
  filtering methods for practitioners. \textit{IEEE Transactions on Signal
  Processing}, \textbf{50}, 736--746, \doi{10.1109/78.984773}.

\bibitem[{Durran and Gingrich(2014)Durran, and Gingrich}]{durran_gingrich_2014}
Durran, D.~R., and M.~Gingrich, 2014: Atmospheric predictability: why
  butterflies are not of practical importance. \textit{Journal of the
  Atmospheric Sciences}, \textbf{71}, 2476–2488,
  \doi{10.1175/JAS-D-14-0007.1}.

\bibitem[{Evensen(1994)}]{evensen_1994}
Evensen, G., 1994: Sequential data assimilation with a nonlinear
  quasi-geostrophic model using {M}onte {C}arlo methods to forecast error
  statistics. \textit{J.\ Geophys.\ Res.}, \textbf{99}, 10\,143--10\,162,
  \doi{10.1029/94JC00572}.

\bibitem[{Fletcher(2010)}]{fletcher_2010}
Fletcher, S.~J., 2010: Mixed {G}aussian-lognormal four-dimensional data
  assimilation. \textit{{Tellus A: Dynamic Meteorology and Oceanography}},
  \textbf{62}, 266--287, \doi{10.1111/j.1600-0870.2009.00439.x}.

\bibitem[{Fletcher et~al.(2023)}]{fletcher_et_al_2023}
Fletcher, S.~J., and Coauthors, 2023: Lognormal and mixed {G}aussian-lognormal
  {K}alman filters. \textit{Mon.\ Wea.\ Rev.}, \textbf{151}, 761–774,
  \doi{0.1175/MWR-D-22-0072.1}.

\bibitem[{Frolov et~al.(2024)}]{frolov_et_al_2024}
Frolov, S., and Coauthors, 2024: Local volume solvers for {Earth} system data
  assimilation: {I}mplementation in the framework for {Joint Effort for Data
  Assimilation Integration}. \textit{Journal of Advances in Modeling Earth
  Systems}, \textbf{16}, e2023MS003\,692, \doi{10.1029/2023MS003692}.

\bibitem[{Gaspari and Cohn(1999)Gaspari, and Cohn}]{gaspari_cohn_1999}
Gaspari, G., and S.~Cohn, 1999: Construction of correlation functions in two
  and three dimensions. \textit{Quart.\ J.\ Roy.\ Meteor.\ Soc.}, \textbf{125},
  723--757, \doi{10.1002/qj.49712555417}.

\bibitem[{Gordon et~al.(1993)Gordon, Salmond,, and Smith}]{gordon_et_al_1993}
Gordon, N.~J., D.~J. Salmond, and A.~F.~M. Smith, 1993: Novel approach to
  nonlinear/non-{G}aussian {B}ayesian state estimation. \textit{Proc. Inst.
  Elect. Eng. F}, \textbf{1400}, 107–113.

\bibitem[{Held et~al.(1995)Held, Pierrehumbert, Garner,, and
  Swanson}]{held_et_al_1995}
Held, I.~M., R.~T. Pierrehumbert, S.~T. Garner, and K.~L. Swanson, 1995:
  Surface quasi-geostrophic dynamics. \textit{J.\ Fluid\ Mech.}, \textbf{282},
  1–20, \doi{10.1017/S0022112095000012}.

\bibitem[{Hu and van Leeuwen(2021)Hu, and van Leeuwen}]{hu_vanLeeuwen_2021}
Hu, C.-C., and P.~J. van Leeuwen, 2021: A particle flow filter for
  high-dimensional system applications. \textit{Quart.\ J.\ Roy.\ Meteor.\
  Soc.}, \textbf{147}, 2352–2374, \doi{10.1002/qj.4028}.

\bibitem[{Hu et~al.(2023)Hu, Dance, Bannister, Chipilski, Guillet, Macpherson,
  Weissmann,, and Yussouf}]{hu_et_al_2023}
Hu, G., S.~L. Dance, R.~N. Bannister, H.~G. Chipilski, O.~Guillet,
  B.~Macpherson, M.~Weissmann, and N.~Yussouf, 2023: Progress, challenges, and
  future steps in data assimilation for convection-permitting numerical weather
  prediction: {Report} on the virtual meeting held on 10 and 12 november 2021.
  \textit{Atmos. Sci. Let.}, \textbf{24}, e1130, \doi{10.1002/asl.1130}.

\bibitem[{Hunt et~al.(2007)Hunt, Kostelich,, and Szunyogh}]{hunt_et_al_2007}
Hunt, B.~R., E.~J. Kostelich, and I.~Szunyogh, 2007: Efficient data
  assimilation for spatiotemporal chaos: {A} local ensemble transform {K}alman
  filter. \textit{Physica D}, \textbf{230}, 112--126,
  \doi{10.1016/j.physd.2006.11.008}.

\bibitem[{Isaksen et~al.(2010)Isaksen, Bonaita, Buizza, Fisher, Haseler,
  Leutbecher,, and Raynaud}]{isaksen_et_al_2010}
Isaksen, L., M.~Bonaita, R.~Buizza, M.~Fisher, J.~Haseler, M.~Leutbecher, and
  L.~Raynaud, 2010: Ensemble of data assimilations at {ECMWF}. \textit{ECMWF
  Technical Memoranda}, \textbf{636}, 1--41.

\bibitem[{Jones et~al.(2010)Jones, Knopfmeier, Wheatley, Creager, Minnis,, and
  Palikonda}]{jones_et_al_2016}
Jones, T.~A., K.~Knopfmeier, D.~Wheatley, G.~Creager, P.~Minnis, and
  R.~Palikonda, 2010: Storm-scale data assimilation and ensemble forecasting
  with the {NSSL} experimental {Warn-on-Forecast} system. {Part II: Combined}
  radar and satellite data experiments. \textit{Wea.\ Forecasting},
  \textbf{30}, 1795–1817, \doi{10.1175/WAF-D-15-0043.1}.

\bibitem[{Kloeden and Platen(1992)Kloeden, and Platen}]{SDE}
Kloeden, P.~E., and E.~Platen, 1992: \textit{Numerical solution of stochastic
  differential equations}, Applications of Mathematics (New York), Vol.~23.
  Springer-Verlag, Berlin, xxxvi+632 pp.

\bibitem[{Luo and Hu(2021)Luo, and Hu}]{DBLP:conf/iccv/LuoH21}
Luo, S., and W.~Hu, 2021: Score-based point cloud denoising. \textit{2021
  {IEEE/CVF} International Conference on Computer Vision, {ICCV} 2021,
  Montreal, QC, Canada, October 10-17, 2021}, {IEEE}, 4563--4572,
  \doi{10.1109/ICCV48922.2021.00454},
  \urlprefix\url{https://doi.org/10.1109/ICCV48922.2021.00454}.

\bibitem[{Milstein(1975)}]{milstein_1975}
Milstein, G.~N., 1975: Approximate integration of stochastic differential
  equations. \textit{Theory of probability \& its applications}, \textbf{19},
  557--562, \doi{10.1137/1119062}.

\bibitem[{Poterjoy(2022)}]{poterjoy_2022}
Poterjoy, J., 2022: Implications of multivariate non-{G}aussian data
  assimilation for multi-scale weather prediction. \textit{Mon.\ Wea.\ Rev.},
  \textbf{150}, 1475--1493, \doi{10.1175/MWR-D-21-0228.1}.

\bibitem[{Poterjoy et~al.(2017)Poterjoy, Sobash,, and
  Anderson}]{poterjoy_et_al_2017}
Poterjoy, J., R.~A. Sobash, and J.~L. Anderson, 2017: Convective-scale data
  assimilation for the {W}eather {R}esearch and {F}orecasting model using the
  local particle filter. \textit{Mon.\ Wea.\ Rev.}, \textbf{145}, 1897–1918,
  \doi{10.1175/MWR-D-16-0298.1}.

\bibitem[{Pulido and van Leeuwen(2019)Pulido, and van
  Leeuwen}]{pulido_vanLeeuwen_2019}
Pulido, M., and P.~J. van Leeuwen, 2019: Sequential {Monte Carlo} with kernel
  embedded mappings: {The} mapping particle filter. \textit{J.\ Comp.\ Phys.},
  \textbf{396}, 400--415, \doi{10.1016/j.jcp.2019.06.060}.

\bibitem[{Rojahn et~al.(2023)Rojahn, Schenk, van Leeuwen,, and
  Potthast}]{rojahn_et_al_2023}
Rojahn, A., N.~Schenk, P.~J. van Leeuwen, and R.~Potthast, 2023: Particle
  filtering and {G}aussian mixtures - on a localized mixture coefficients
  particle filter ({LMCPF}) for global {NWP}. \textit{J.\ Meteor.\ Soc.\
  Japan}, \textbf{101}, 233--253, \doi{10.2151/jmsj.2023-015}.

\bibitem[{Rotunno and Snyder(2008)Rotunno, and Snyder}]{rotunno_snyder_2008}
Rotunno, R., and C.~Snyder, 2008: A generalization of {L}orenz's model for the
  predictability of flows with many sclaes of motion. \textit{J.\ Atmos.\
  Sci.}, \textbf{65}, 1063–1076, \doi{10.1175/2007JAS2449.1}.

\bibitem[{Schraff et~al.(2016)Schraff, Reich, Rhodin, Schomburg, Stephan,
  Periáñez,, and Potthast}]{schraff_et_al_2016}
Schraff, C., H.~Reich, A.~Rhodin, A.~Schomburg, K.~Stephan, A.~Periáñez, and
  R.~Potthast, 2016: Kilometre-scale ensemble data assimilation for the {COSMO}
  model ({KENDA}). \textit{Quart.\ J.\ Roy.\ Meteor.\ Soc.}, \textbf{142},
  1453--1472, \doi{10.1002/qj.2748}.

\bibitem[{Smith et~al.(2023)Smith, Penny, Platt,, and Chen}]{smith_et_al_2023}
Smith, T.~A., S.~G. Penny, J.~A. Platt, and T.-C. Chen, 2023: Temporal
  subsampling diminishes small spatial scales in recurrent neural network
  emulators of geophysical turbulence. \textit{Journal of Advances in Modeling
  Earth Systems}, \textbf{15}, e2023MS003\,792, \doi{10.1029/2023MS003792}.

\bibitem[{Song et~al.(2021)Song, Sohl-Dickstein, Kingma, Kumar, Ermon,, and
  Poole}]{song2021scorebased}
Song, Y., J.~Sohl-Dickstein, D.~P. Kingma, A.~Kumar, S.~Ermon, and B.~Poole,
  2021: Score-based generative modeling through stochastic differential
  equations. \textit{International Conference on Learning Representations},
  \urlprefix\url{https://openreview.net/forum?id=PxTIG12RRHS}.

\bibitem[{Spantini et~al.(2022)Spantini, Baptista,, and
  Marzouk}]{spantini_et_al_2022}
Spantini, A., R.~Baptista, and Y.~Marzouk, 2022: Coupling techniques for
  nonlinear ensemble filtering. \textit{SIAM Review}, \textbf{144}, 921--953,
  \doi{10.1137/20M1312204}.

\bibitem[{T{\"o}dter et~al.(2016)T{\"o}dter, Kirchgessner, Nerger,, and
  Ahrens}]{todter_et_al_2016}
T{\"o}dter, J., P.~Kirchgessner, L.~Nerger, and B.~Ahrens, 2016: Assessment of
  a nonlinear ensemble transform filter for high-dimensional data assimilation.
  \textit{Mon.\ Wea.\ Rev.}, \textbf{144}, 409--427,
  \doi{10.1175/MWR-D-15-0073.1}.

\bibitem[{Tulloch and Smith(2009{\natexlab{a}})Tulloch, and
  Smith}]{tulloch_smith_2009b}
Tulloch, R., and K.~S. Smith, 2009{\natexlab{a}}: A note on the numerical
  presentation of surface dynamics in quasigeostrophic turbulence. \textit{J.\
  Atmos.\ Sci.}, \textbf{66}, 1063--1068, \doi{10.1175/2008JAS2921.1}.

\bibitem[{Tulloch and Smith(2009{\natexlab{b}})Tulloch, and
  Smith}]{tulloch_smith_2009a}
Tulloch, R., and K.~S. Smith, 2009{\natexlab{b}}: Quasigeostrophic turbulence
  with explicit surface dynamics: {Application} to the atmospheric energy
  spectrum. \textit{J.\ Atmos.\ Sci.}, \textbf{66}, 450--467,
  \doi{10.1175/2008JAS2653.1}.

\bibitem[{van Leeuwen(2009)}]{vanLeeuwen_2009}
van Leeuwen, P.~J., 2009: Particle filtering in geophysical systems.
  \textit{Mon.\ Wea.\ Rev.}, \textbf{137}, 4089–4114,
  \doi{10.1175/2009MWR2835.1}.

\bibitem[{van Leeuwen et~al.(2019)van Leeuwen, K{\"u}nsch, Nerger, Potthast,,
  and Reich}]{vanLeeuwen_et_al_2019}
van Leeuwen, P.~J., H.~R. K{\"u}nsch, L.~Nerger, R.~Potthast, and S.~Reich,
  2019: Particle filters for high-dimensional geoscience applications: a
  review. \textit{Quart.\ J.\ Roy.\ Meteor.\ Soc.}, \textbf{145}, 2335--2365,
  \doi{10.1002/qj.3551}.

\bibitem[{Wang et~al.(2021)Wang, Chipilski, Bishop, Satterfield, Baker,, and
  Whitaker}]{wang_et_al_2021}
Wang, X., H.~G. Chipilski, C.~H. Bishop, E.~Satterfield, N.~Baker, and J.~S.
  Whitaker, 2021: A multiscale local gain form ensemble transform kalman filter
  ({MLGETKF}). \textit{Mon.\ Wea.\ Rev.}, \textbf{149}, 605–622,
  \doi{10.1175/MWR-D-20-0290.1}.

\bibitem[{Whitaker and Hamill(2012)Whitaker, and Hamill}]{whitaker_hamill_2012}
Whitaker, J.~S., and T.~Hamill, 2012: Evaluating methods to account for system
  rrrors in ensemble data assimilation. \textit{Mon.\ Wea.\ Rev.},
  \textbf{140}, 3078–3089, \doi{10.1175/MWR-D-11-00276.1}.

\end{thebibliography}
